\documentclass[aip,reprint]{revtex4-1}

\usepackage{amsmath, amssymb}
\usepackage{xcolor}
\usepackage{verbatim}
\usepackage{graphicx}
\usepackage{natbib}
\usepackage{siunitx}
\usepackage{upgreek}
\usepackage[textsize=tiny]{todonotes}
\usepackage{longtable}

\usepackage[version=4,arrows=pgf-filled,
textfontname=sffamily,
mathfontname=mathsf]{mhchem}

\draft % marks overfull lines with a black rule on the right

\begin{document}

%\preprint{}

\title{Taming polymorphism of tubule self-assembly using templated growth}



\author{Sirui Liu}
\affiliation{Martin A. Fisher School of Physics, Brandeis University, Waltham, MA, 02453, USA}

\author{Thomas E. Videb\ae k}
\affiliation{Martin A. Fisher School of Physics, Brandeis University, Waltham, MA, 02453, USA}

\author{W. Benjamin Rogers}
\email{wrogers@brandeis.edu}
\affiliation{Martin A. Fisher School of Physics, Brandeis University, Waltham, MA, 02453, USA}


%\date{\today}

\begin{abstract}
Self-closing assembly is prone to polymorphism due to thermally-excited bending fluctuations, which permit the formation of off-target assemblies at the point of self-closure. One way to overcome this source of polymorphism is to use templated growth, a process in which assembly initiates from a precisely-defined seed rather than by spontaneous nucleation. We explore this approach to quelling polymorphism in the self-closing assembly of cylindrical tubules assembled from DNA-origami subunits with user-specified inter-subunit binding angles and specific interactions. We develop two strategies to create seeds with precisely-defined diameters and helicity: 1) using multicomponent assembly; and 2) purifying a specific seed-type from a polymorphic mixture using gel electrophoresis and gel extraction. By tuning the seed and monomer concentrations, and adjusting the assembly temperature, we determine the conditions under which tubules grow from the seed while avoiding spontaneous nucleation. We observe that templated tubules tend to follow the guidance of the seed, thereby increasing the selectivity of the target geometry. Also, we find that by tuning the diameter of the seed, one can template the growth of monodisperse tubules over a range of target diameters, even while using a single monomer type with a single preferred local curvature. Our results demonstrate that employing precisely defined seeds to guide assembly can significantly decrease polymorphism in self-closing assembly in a controllable and economical way.

\end{abstract}


\maketitle %\maketitle must follow title, authors, abstract and \pacs

\section*{Introduction}

%\textbf{ Paragraph 1 will introduce self-closing assembly, provide examples of self-closing assemblies in nature, and introduce polymorphism in self-closing assembly. }

Self-closing assembly, in which the global size and geometry of the target structure is encoded in the local geometry and interactions of the subunits, is an economic strategy for synthesizing self-limiting structures with one or more finite-size dimensions~\cite{hagan2021equilibrium,jacobs2025assembly}. Due to its economy, self-closing assembly is ubiquitous in biological systems, with examples including microtubules~\cite{nogales1999high,desai1997microtubule}, viral capsids~\cite{caspar1962physical,garmann2016physical}, and cell membranes~\cite{deuling1976curvature}. But this economy of design is both a blessing and a curse. Because many self-closing structures have high degrees of symmetry, they are susceptible to polymorphism, which poses a crucial challenge to both biological and synthetic systems~\cite{chretien1991new,nguyen2009invariant, mohajerani2022multiscale, helfrich1986size, Hayakawa2022,videbaek2024economical, wagenbauer2017gigadalton}.

%\textbf{Paragraph 2 will introduce seeded/templated  growth as a strategy to minimize polymorphism. Introduce biological and synthetic examples.}

Templated growth, an assembly pathway along which the target structure grows from a precisely defined seed, provides a potential route to circumvent this type of polymorphism. A canonical example in Biology is the self-assembly of microtubules, structural filaments that grow in the cell cytoplasm. In vitro, purified microtubule proteins can self-assemble to form tubules, but do so with a distribution of protofilament numbers, resulting in tubules with different diameters and helicities~\cite{chretien1991new}. Interestingly, in vivo, microtubules predominantly form with thirteen protofilaments, a feature which has been attributed to several secondary protein structures, one of which is the gamma-tubulin ring complex, a multi-protein assembly that serves to nucleate microtubules~\cite{bohm1984effect, roostalu2017microtubule}. In a similar fashion, researchers have used templated growth to control the diameter of nanotubules made of from synthetic building blocks, such as DNA tiles. By creating a small cylindrical seed using DNA origami, nanotubules of a specific size could be targeted~\cite{jorgenson2017self,mohammed2013directing}. Unlike the example of microtubules, however, this synthetic seed is made from a single subunit. This strategy has also been observed in the self-assembly of large viral capsids, in which the initial formation of a protocapsid core templates the assembly of the final capsid on its surface, before subsequent disassembly and extraction of the protocapsid~\cite{greene1994binding, thuman1996three}. While synthetic systems have successfully templated the assembly of closed shells from solid seeds, such as gold nanoparticles~\cite{chen2006nanoparticle} or DNA origami~\cite{seitz2023dna}, the seeds in those cases were again not multicomponent assemblies, in contrast to the protocapsid example. 

%\textbf{Paragraph 3: The "In this paper, we ..." paragraph that summarizes the major results of this project.}

\begin{figure*}
    \centering
    \includegraphics[width=\linewidth]{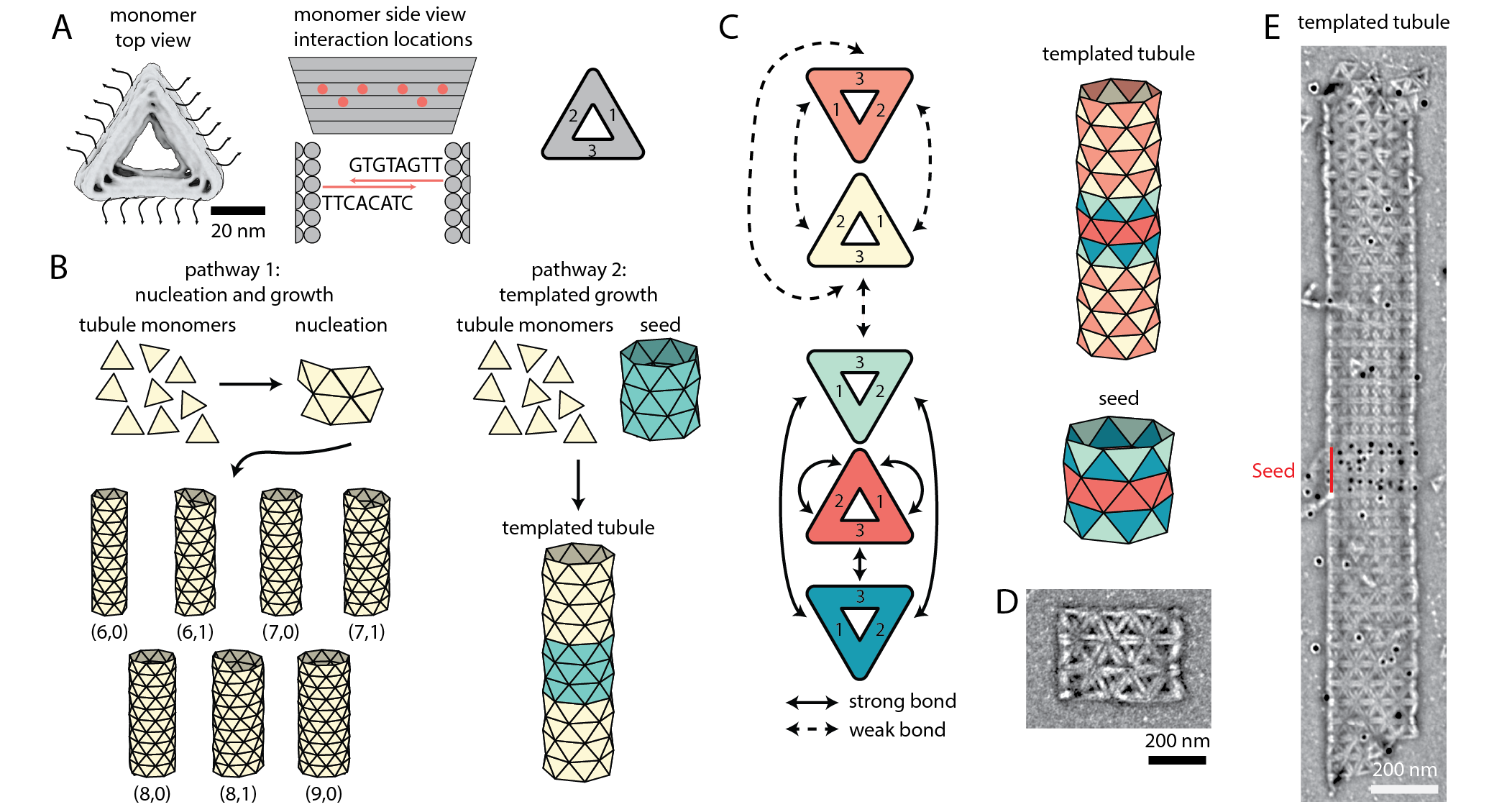}
    \caption{ \textbf{Assembling self-closing tubules via templated growth}. (A) cryo-EM single-particle reconstruction and schematics of our DNA-origami triangular monomer. Six single-stranded DNA handles extend from specific locations of each face (red circles) and mediate specific interactions via base pairing. (B) Assembly can proceed along two pathways: (1) nucleation and growth; and (2) templated growth from a seed.  (C) Specific inter-subunit interactions needed to assemble a tubule and a seed. Lines indicate favorable interactions. Solid lines represent stronger interactions and dashed lines indicated weaker interactions. (D) TEM image of an assembled seed (Fig.~S19 shows the widefield of seed assembly). (E) TEM image of a templated tubule. The presence of DNA-conjugated gold nanoparticles (black dots) in the middle of the tubule indicates the location of the seed. }
    \label{Fig:fig-intro}
\end{figure*}

In this paper, we explore templated growth as a strategy to tame polymorphism in the self-assembly of cylindrical tubules, mimicking the in-vivo assembly of microtubules, by self-assembling a seed from many components. We determine the conditions under which it is possible to grow cylindrical tubules from achiral, ring-shaped seeds while avoiding spontaneous nucleation through the careful control of the initial supersaturation and the ratio of seeds to assembling monomers. We create ring-shaped seeds with controlled initial diameters using two strategies: 1) using multicomponent assembly to design and synthesize seeds of user-specified diameters; and 2) using gel extraction to physically purify seeds of a specific diameter from an initially polymorphic mixture. Importantly, in both cases, we observe that tubules are strongly biased to form tubules that are isostructural to the seed, for a variety of seed diameters, even when the seed diameter is different from the preferred curvature of the monomers. Taken together, these results demonstrate that it is possible to use kinetic control to create monodisperse, self-closing assemblies from a simple pool of monomers by controlling the nucleation and growth pathways to assembly.

\section*{Results}

\subsection*{Our experimental system}

%\textbf{Paragraph 4 will introduce our experimental system: DNA origami triangles with specific edge-edge interactions and programmed curvature.}

We use the molecular engineering capabilities of DNA origami to create triangular colloidal particles with geometrically-programmed, specific interactions. In particular, we synthesize equilateral triangular building blocks that are roughly 50-nm in edge-length. We program the local curvature by beveling the three edges of the triangular particle independently~\cite{videbaek2024economical}. This preferred local curvature drives a growing assembly to close upon itself at a specified diameter and pitch, thereby forming a cylindrical tubule~\cite{fang2022polymorphic}. We also encode specific edge-to-edge interactions between subunits by designing single-stranded DNA that extends from the edges of the particles (Fig.~\ref{Fig:fig-intro}A). Throughout this work, we create different subunit species by changing the sequences of the edge interaction while maintaining the same monomer geometry. We use a pair of integers ($m,n$) to classify the cylindrical tubules that result from self-assembly. Together, the two integer values define the minimal loop that traverses triangle edges: $m$ denotes steps in the circumferential direction and $n$ denotes steps in the axial direction. The monomer we design has beveled edges targeting a (6,0) achiral tubule.

%\textbf{Paragraph 5 introduces the assembly experiment, explaining unseeded assembly and how it results in a broad distribution of tubule types, as well as how we characterize them and their distribution.(Fig.\ref{Fig:fig-intro}A-Left)}

In the absence of a seed, we find that tubules do indeed assemble but form a polymorphic mixture of many distinct tubule geometries. We test the assembly of our subunit by performing experiments at a concentration of 10~nM monomers and 20 mM MgCl$_2$, and incubating at 36~\textdegree C for two weeks. Post assembly, we characterize the structures that form using negative-stain transmission electron microscopy (TEM) (Fig.~S10), classifying a zoo a tubule geometries with slightly different diameters and helicities. We attribute this structural variability to an assembly pathway that proceeds via  nucleation and growth (Fig.~\ref{Fig:fig-intro}B). Since tubule assemblies begin as an unclosed curved sheet, fluctuations of the sheet curvature allow the sheet to close in different configurations~\cite{Hayakawa2022,fang2022polymorphic,videbaek2024economical}. Because this source of polymorphism is intrinsic to the rigidity of the inter-subunit contacts~\cite{videbaek2025measuring}, the formation of self-closing structures through spontaneous nucleation is always susceptible to this type of polymorphism.

\subsection*{Templated assembly: The basics}

\begin{figure*}
    \centering
    \includegraphics[width=\linewidth]{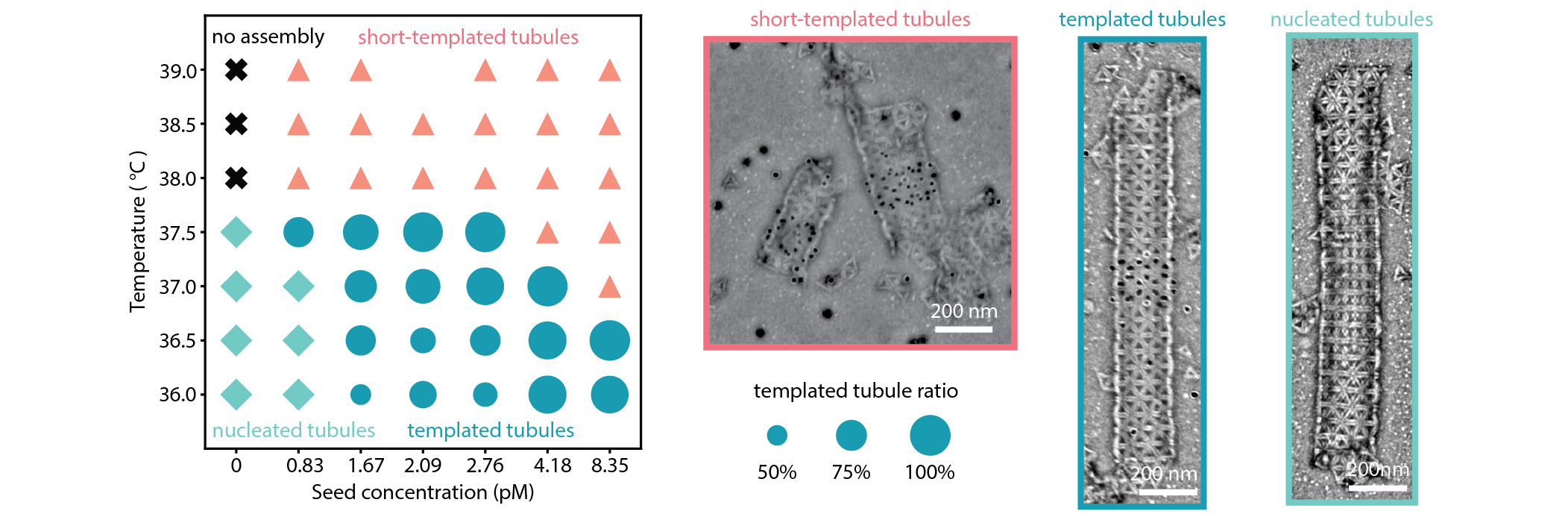}
    \caption{\textbf{Templated assembly state diagram}. Each data point represents an assembly outcome from experiments. We define the templated tubule state as an assembly with an average tubule length greater than 700~nm in which more than 50\% of the tubules grow from a seed (Fig.~S6 shows the state diagram for different cutoff lengths). The size of the circle is proportional to the fraction of templated tubules, which is the number of templated tubules divided by the total number of tubules. The micrographs show representative structures for various kinetic regimes of the state diagram.}
    \label{Fig:kinetic_study}
\end{figure*}

%\textbf{ Paragraph 6 discusses how to make a simple seeded tubule (Fig. \ref{Fig:fig-intro}B) and show a sample of the TEM image (Fig.\ref{Fig:fig-intro}C).}

To overcome the polymorphism that results from spontaneous nucleation and self-closure, we pursue an alternate strategy: templated growth. We first design a seed. The design of the seed has three general requirements: 1) a seed should be finite in size; 2) it should be stable; and 3) it should be built from subunits with interactions that are stronger than the interactions of the monomers that will grow the tubules from the initial seeds. To meet these requirements, we design our seed to be an achiral, finite-length tubule made of three layers, which we realize using the interactions shown in Fig.~\ref{Fig:fig-intro}C. To make the seed monomers and tubule monomers distinguishable, the center of the seed monomers has single-stranded DNA binding sites that can bind DNA-coated gold nanoparticles. The exposed edge of the seed can then interact with the tubule monomers, allowing for tubule growth. To ensure that the intra-seed interactions are stronger than the interactions of the tubule-growing monomers, we use 6-base-pair-long (bp) binding domains for the tubule interactions and 7-bp-long binding domains for the seed interactions. 

Assembly experiments demonstrate the ability to assemble finite-length seeds, to selectively label them with DNA-functionalized gold nanoparticles, and to use those seeds to template the growth of long, single-domain cylindrical tubules. We first assemble seeds by mixing seed monomers with equal stoichiometry and incubating the sample at 40~\textdegree C in 20~mM MgCl$_2$ for four days. The TEM image in Fig.~\ref{Fig:fig-intro}D shows that the seed assembles as designed. We then mix the tubule monomers with the pre-assembled seed and incubate the solution at 37~\textdegree C in 20~mM MgCl$_2$ for two weeks. Before observing the assembly with TEM, we add gold nanoparticles that bind specifically to the seed monomers to distinguish templated and spontaneously-nucleated tubules. Under TEM we find both types of tubules (Fig.~\ref{Fig:fig-intro}E and Fig.~S4). Despite the pre-assembled seeds' ability to template growth, spontaneous nucleation can still occur, indicating that we need to develop a better understanding of the assembly kinetics to select conditions in which templated growth is favorable yet spontaneous nucleation is rare.

\subsection*{Kinetic control}

%\textbf{Paragraph 8 describes and demonstrates in experiment the kinetic control that's required for seeded growth (Fig 2). The subunit-subunit interactions need to be strong enough that the seed will grow but not so strong that the subunits will nucleate new tubules on their own. We control these rates by tuning the temperature and seeded monomer ratio.}

We first characterize how the assembly parameters impact the kinetics of tubule formation to identify combinations of parameter values that bias assembly toward templated-growth pathways. In particular, we prepare mixtures with 10~nM of tubule monomers together with pre-assembled seeds, whose concentrations vary from 0~pM to 8.35~pM. Since DNA hybridization, and therefore the nucleation and growth rates, is sensitive to the temperature, we incubate  assemblies over a range of temperatures. For each condition, we image the samples using TEM and attempt to classify at least 100 assembled structures. This classification is possible when tubules form, and we can identify each assembly as either a templated or nucleated structure based on the presence of a seed. Under conditions at which assembly is kinetically arrested or does not yield well-assembled tubules, we look over at least ten grid squares and classify as many tubules as are present (see Fig.~S5A for counts of assemblies). For reference, when assembled tubules are observed, we typically find 100 assemblies on five grid squares. 

We find a variety of behaviors and group the associated assembly outcomes into four different kinetic regimes: nucleated tubules, templated tubules, short-templated tubules, and no assembly. The distinction between short-templated tubules and templated tubules is made to emphasize the growth rate: we hypothesize that short-templated tubules have a slower characteristic growth rate than templated tubules. To this end, we place a cutoff on the length of tubules at 700~nm to separate these slower and faster growing assemblies (Fig.~S5B shows the average lengths of tubules). The different kinetic regimes are summarized in a state diagram in Fig.~\ref{Fig:kinetic_study}, with each point showing the majority outcome.

This state diagram informs us about the role of seeds and the suppression of spontaneous nucleation. In the absence of seeds (i.e., a seed concentration of 0~pM), we find spontaneously nucleated tubules at low temperatures and no assembly at high temperatures. At low seed concentration and low assembly temperatures, we find that assembly is similarly dominated by spontaneous nucleation. Upon increasing the seed concentration, the pathway to assembly switches from spontaneous nucleation to templated growth, with the vast majority of structures being grown from a seed above a seed concentration of 1.67~pM. However, if the seed concentration becomes too high, the tubules that form become shorter and shorter, presumably because there is an excess of nucleation sites that causes monomers to be distributed into many assemblies. At intermediate values of the seed concentration (i.e., between 1.7--2.8~pM), we find that the fraction of templated tubules increases upon increasing temperature, before eventually transitioning to the short-templated tubule regime, indicating slow growth rates. We attribute the smaller fraction of templated tubules at lower temperatures to an increase in the spontaneous nucleation rate. 

We use our state diagram to inform our subsequent assembly experiments. We aim to create an assembly at a temperature high enough to have an appreciably high free-energy barrier that suppresses spontaneous nucleation, but not so high as to suppress tubule growth~\cite{mohammed2013directing, minev2021robust, zhang2020programming}. In addition, the concentration of templated monomers should not be too high, as this can deplete free monomers and shorten the average tubule length. Hence, we select 37~\textdegree C and   2.8~pM seed concentration for the (1,0) PU cell as our assembly condition. Under these conditions, we find that 90\% of assemblies in solution are templated tubules and the tubules are much longer than the seeds.

Having found an optimal growth condition for templated assembly, we now compare the polymorphism of templated and spontaneously-nucleated tubules. We use the simplest seed possible, which is three-layers long and composed of three distinct particle species (Fig.~\ref{fig_multispecies_seed}A). To characterize the resultant polymorphism, we measure the widths of the templated tubules using TEM. If a tubule is polymorphic within a single structure (Fig.~S7), we measure the width of each section. We first characterize the assembly of tubules due to spontaneous nucleation and plot their widths (Fig.~\ref{fig_multispecies_seed}B), finding a broad distribution, as mentioned above. We then characterize both the initial distribution of the seed diameters before mixing with the tubule monomers (Fig~\ref{fig_multispecies_seed}C, red points) and the resultant templated tubules (Fig.~\ref{fig_multispecies_seed}C, bars). While the templated assembly has a marginally higher yield for widths that match the seed width, the two distributions are quite similar. One important reason is that the seeds that we self-assembly are similarly polymorphic and have a distribution similar to the spontaneously-nucleated tubules. 

\subsection*{Making precise seeds with assembly complexity}

\begin{figure}[!t]
    \centering
    \includegraphics[width=\linewidth]{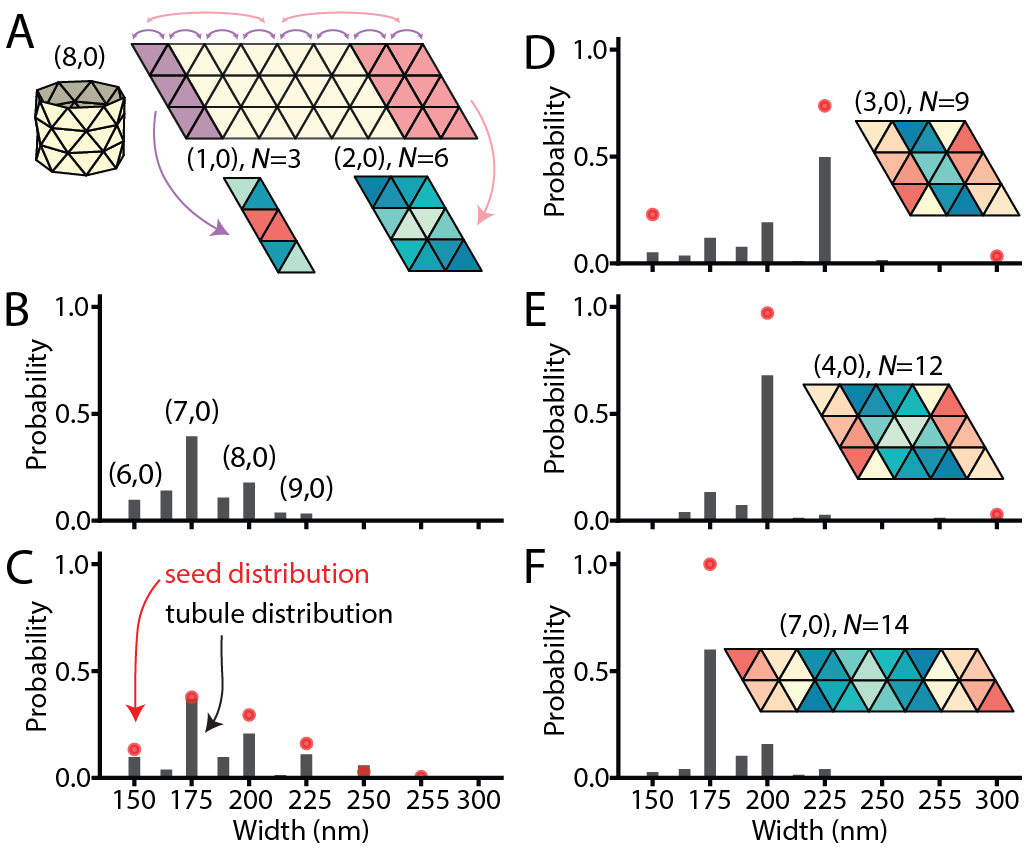}
    \caption{\textbf{Using multispecies tilings to assemble specific seeds.}
    (A) Multispecies tilings can be used to define the seed length and restrict the allowed seed states. A seed can  be conceptualized of as a segment of a planar tiling with periodic boundary conditions. We break up the seed into parallelograms with different widths (the purple and red shaded regions); these parallelograms are possible principle unit (PU) cells. Only when the width of the PU cell is a divisor of the circumference is the formation of a seed with that circumference allowed. Once a size of PU cell is chosen, we color in different triangles to identify the number of species required to assemble it~\cite{hayakawa2024symmetry}. (B) Tubule width probability distribution from spontaneously nucleated tubules in the absence of seeds. (C--F) Tubule width probability distributions for assemblies with multispecies seeds. The red points show the distribution of the assembled seeds, and the gray bars represent the probability of the assembled tubules. The inset figures show the associated PU cell: (C) (1,0) PU cell, (D) (3,0) PU cell, (E) (4,0) PU cell, and (F) (7,0) PU cell. All tubule assemblies were grown for two weeks. All templated assemblies have 0.33~nM of seed-monomers, which corresponds to estimated seed concentrations of  2.8~pM (C), 1.64~pM (D) , 0.6~pM (E), and  1.62~pM (F).
   }
    \label{fig_multispecies_seed}
\end{figure}

To overcome the seed-diameter dispersity, we pursue two conceptually distinct approaches to making highly specific seeds. First, we explore the possibility of using multi-component assembly to make less polymorphic seeds by eliminating nearby off-target seed-sizes~\cite{videbaek2024economical}. Then, we pursue using gel electrophoresis and gel extraction to purify a specific seed from a polymorphic mixture by physical separation.

One approach to creating a more specific seed is to increase the number of distinct monomer species from which the seeds are assembled~\cite{hayakawa2024symmetry,videbaek2024economical}. Just as for the tubules themselves, thermal fluctuations in the inter-subunit binding angles allow the seed-forming monomers to assemble into seeds with varying diameters if the interaction matrix allows for it. Therefore, because the smallest repeating building block of the seed is a one-monomer-wide strip, any seed with an integer value of $m$ and $n=0$ can in principle form (Fig.~\ref{fig_multispecies_seed}A). We call this strip the (1,0) primitive unit cell (PU cell).

By increasing the size of the primitive unit cell, we can restrict the number of geometrically allowed configurations of the seeds, thereby enhancing their selectivity. We utilize the inverse design rules from Hayakawa et al.~\cite{hayakawa2024symmetry} to design the interactions for (3,0), (4,0), and (7,0) PU cells (Fig.~\ref{fig_multispecies_seed}D--F). For these different seed designs, the number of unique subunit species grows as the product of the $m$ number of the PU cell and the number of layers in the seed. As a result, we design the (7,0) PU cell to only be two layers tall to reduce the number of subunit species required.

%\textbf{Paragraph 11 will present the experimental result of assembling a multispecies seed to show that it constrains the allowed seed types and therefore alters the seed distribution.(Fig.\ref{fig_multispecies_seed}B-F red dots)}

We assemble the higher $m$ number PU cells and quantify the distribution of seed diameters for each design (Fig.~\ref{fig_multispecies_seed}D–-F, red dots). We find that increasing the size of the PU cell narrows the spectrum of accessible seeds, as expected. While the (4,0) and (7,0) PU cell seeds show nearly complete selectivity, the (3,0) PU cell assemblies form predominantly into two seed types (Fig.~\ref{fig_multispecies_seed}D). Because the spontaneously-nucleated tubule distribution has the highest probability of assembly in the (7,0) and (8,0) tubule configurations, we expect that the (4,0) and (7,0) PU cell seeds will effectively template the growth of tubules. In contrast, the (3,0) PU cell selects for states that are not geometrically preferred by the monomer, leading to the formation of multiple seed types. While these experiments show that changing assembly complexity provides some ability to tune the seed diameter, this approach has some limitations owing that arise from the fixed geometry of the underlying monomer and the flexibility of the inter-monomer contacts.

We observe that higher complexity seeds show a greater ability to reduce polymorphism in the resultant tubule distribution. The tubule distributions from the (3,0), (4,0), and (7,0) PU cell seeds all show a single dominant population of tubules that significantly more pronounced than for the case of nucleated tubules (Fig.~\ref{fig_multispecies_seed}D--F). Also, these peak-widths match the widths of the initial seed distributions used to template growth, and TEM images confirm that the growing tubules follow the guidance of the seed (Fig.~S8). Interestingly, not only do these results demonstrate how to increase the yield of a specific tubule state, but that it is possible to tune the width distribution of tubules simply by changing the seed circumference, without needing to use different monomer geometries.

Despite the ability for seeds to successfully template growth, we still observe tubules with widths that are not present in the seed distribution. Of course, some of these arise from spontaneous nucleation, but we also find two other forms of polymorphism. The first is that tubules can change their tubule-type during elongation (Fig.~S7). This effect seems to be related to the rate of assembly, with more frequent type changes occurring at lower assembly temperatures. Despite conducting assemblies at higher temperatures where growth is slow, we have not been able to find a suitable assembly temperature that eliminates this form of polymorphism. The second source of polymorphism occurs from seed assemblies that fail to close and remain as an open sheet. These sheets still have edges that allow for tubule growth, but they do not force any particular tubule state to form. Occasionally, we find structures that appear to be templated tubules, but with a tubule type that is disallowed from the closed seed (Fig.~S9). This failure mode appears to be more prevalent at higher seed complexity, though it is still fairly rare under those conditions.

\subsection*{Making precise seeds through physical purification}

There is a fundamental limitation to making seeds with multiple species. The seed assembly needs an ever-increasing number of species to target tubules with increasing diameter, and the higher complexity can lead to unintended failure modes~\cite{videbaek2022tiling,videbaek2024economical}. Furthermore, this approach becomes economically intractable and ultimately infeasible once the number of required low-crosstalk interactions becomes too large~\cite{wu2012polygamous}. In this section, we present a completely different approach to creating a precise seed by isolating seeds with specific geometry from a polymorphic mixture. While these results clearly demonstrate feasibility, there is certainly still room to refinement.

\begin{figure}[!t]
    \centering 
    \includegraphics[width=\linewidth]{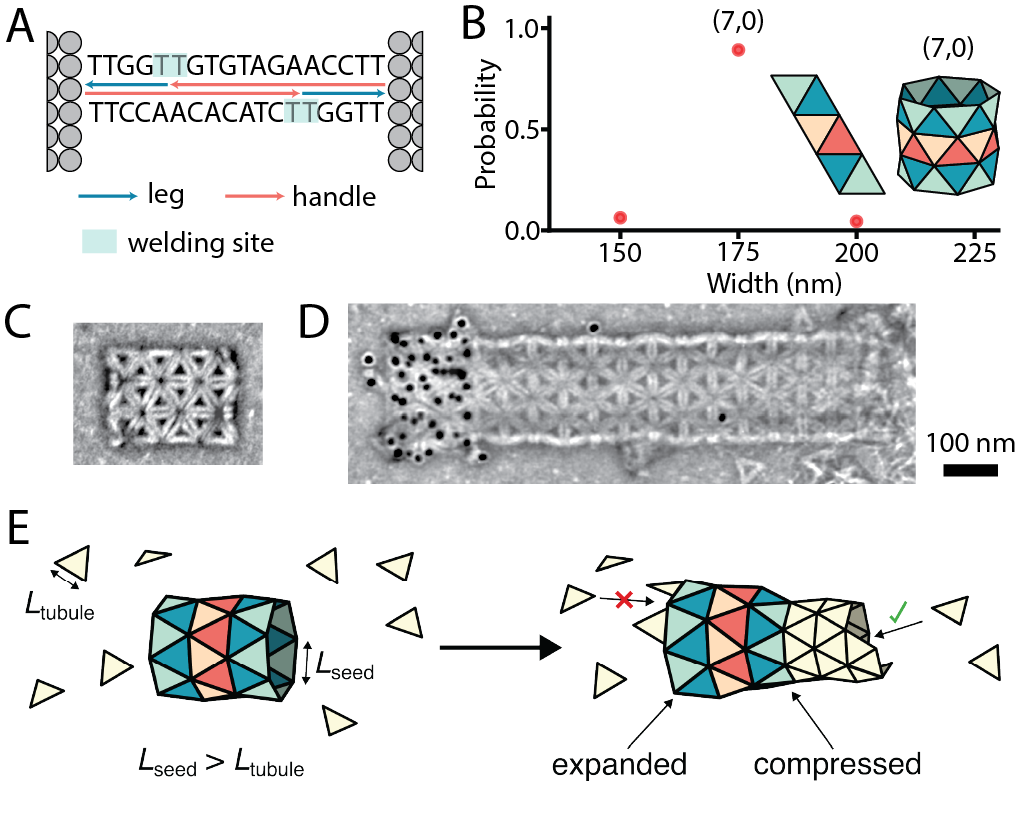}
    \caption{\textbf{Welding and purifying seeds from a polymorphic mixture} (A) UV-weldable monomer handle and leg designs. Shaded blue regions show welding sites. (B) Width distribution of the UV-welded seed purified from the (7,0) band using gel electrophoresis. Insets are the PU cell and the most probability seed geometry. Due to the stronger interactions of the seed monomers, the seed has four particle species to avoid any homophilic interactions. (C) TEM image of a purified (7,0) seed. (D) A templated tubule grown from a purified (7,0) seed. (C) and (D) share the same scale bar. (E) Sketch of how a lattice mismatch between the tubule monomers and the seed monomers could cause polar assembly.}
    \label{Fig:purification}
\end{figure}

Our second approach aims to purify seeds by separating them using gel electrophoresis and gel extraction. Because these processing steps are rather harsh, and seed disassembly is all too likely, we preserve the structure of the seeds by redesigning their interactions to enable the covalent binding of the subunit-subunit contacts (Fig.~\ref{Fig:purification}A). More specifically, we add a section of double-stranded DNA between the binding domain and the body of the DNA-origami monomer, which we call a ``leg''. We also design the sequences such that, in the bound configuration, a pair of thymine bases flank the nick location between the leg and the binding domain. With this pair of thymines in place, we can promote the formation of a cyclobutane pyrimidine dimer between the thymine pair upon exposure to UV light. This technique, called ``UV welding'', has been used previously to stabilize folded DNA-origami structures~\cite{gerling2018sequence}, as well as assemblies of subunits~\cite{monferrer2023dna}. We find that this technique works well to preserve our assembled seeds (Fig.~S1). 

With the added stability from UV welding, we find that our seeds can be purified. We first assemble seeds with the (1,0) PU cell and then UV-weld the structures, finding a polymorphic mixture of seeds. We then run the seed solution through an agarose gel and excise the (7,0) seed band, which we then purify by passing the gel through a DNA gel extraction spin column, separating out the agarose. The seed solution is then concentrated through ultrafiltration (see Methods). Figure~\ref{Fig:purification}B shows the distribution of seeds purified from the (7,0) band of the gel. We find that nearly all seeds are of the correct type (Fig.~\ref{Fig:purification}C), with small amounts of seeds from the neighboring bands.

In some respects, the tubules grown from UV-welded seeds behave similarly to those grown from multispecies seeds, but they also bring with them new surprises (Fig.~\ref{Fig:purification}). Similar to before, the templated assemblies seem to follow the guidance of the seed and show the same penchant for changing their tubule type along the growth pathway. But unlike before, we find that the vast majority of the tubules grown from UV-welded seeds are polar, growing off a single edge of the seed (Fig.~\ref{Fig:purification}D). We attribute this new behavior to subtle differences in the interaction regions at the interface between the seed and tubule monomers. In particular, the tubule monomers have DNA handles with 6-base-long binding domains and 4 unpaired Ts as a spacer. The UV-welded seeds, however, have handles with 14 paired bases and 4 unpaired Ts, leading to a mismatch in bond lengths between the two monomer types of about 3~nm. This mismatch leads to a difference in lattice spacing between the seed and the growing tubule. As a result, we hypothesize that the seed deforms upon tubule elongation (Fig.~\ref{Fig:purification}E), and that this deformation may be enough to inhibit the growth of a tubule from the other side, producing an interesting type of spontaneous symmetry breaking.

While our preliminary results show promise, there are several aspects of our UV-welding approach that could be improved. While we were able to get monomers to attach to the purified seed, this required iterative testing of the DNA sequences that mediate interactions between the seed and the tubule (Fig.~S2). A better understanding of how unbound single-stranded DNA is altered by UV light could lead to better choices for the design of the handle sequences. Another factor that may be responsible for lower monomer attachment could be the lattice spacing mismatch. Therefore, intentionally designing the length of the interaction region to be the same could promote better binding. Lastly, the yield of gel-purified structures is quite low. Partially, this relatively low yield is due to the natural loss of material to various filters, but also, for larger structures, it seems as if many assemblies never enter the gel during electrophoresis. While gel electrophoresis is more approachable, size exclusion chromatography could be a better technique to explore going forward.

\section*{Discussion}

%\textbf{Paragraph 19 should compare and contrast this kinetically controlled route to controlling polymorphism to Thomas' thermodynamically controlled route. What are the similarities/differences? What are the pros/cons?}

In this work, we showed how templated assembly can be used to kinetically control the width distribution of cylindrical tubules by eliminating self-closure. This approach is complementary to other approaches to limiting polymorphism in self-closing assembly, such as the use of complexity to restrict the allowed tubule states~\cite{videbaek2024economical}. Indeed, both of these schemes come with benefits and drawbacks. While using complexity to fully define the allowed states prevents the rise of polymorphism at all points of assembly, it requires a large number of unique subunits, which scales with the square of the tubule circumference~\cite{videbaek2024economical}. Large numbers of subunits can become problematic due to design constraints of creating many low-crosstalk interactions, as well as the slowdown in growth timescales. Defining a seed, especially if it can be purified to a single state, can be done with a small number of components and also allows for control over nucleation, which opens the possibility of tuning the final length of the tubules~\cite{phillips2012physical}. However, templated growth can still allow for polymorphism as the tubule grows.

At first glance, this work resembles previous studies on templated assembly of nanotubules using DNA tiles~\cite{mohammed2013directing,jorgenson2017self}, but several interesting differences warrant remark. The first is the circumference of the tubules that are targeted. Due to the size of the monomers, the tubules in this work are an order of magnitude larger. While DNA origami could be used to seed the growth of DNA-tile nanotubules, we require an initial self-assembly step to produce the seed itself. Additionally, the form of polymorphism exhibited by the DNA tile system is quite different; instead of allowed states being prevalent at both smaller and larger diameters, the smallest diameter tubule is typically preferred~\cite{zhang2019programming}. An interesting feature, while not directly reported on, seems to be that DNA tiles have a low propensity for exhibiting polymorphism within a single nanotubule. For our triangular subunits, we believe that intra-tubule polymorphism arises when growth in the axial direction occurs faster than in the circumferential direction, allowing for secondary closure events to occur~\cite{fang2022polymorphic}. For DNA tiles, monomers typically require two binding sites to attach to a structure~\cite{rothemund2004design}. This feature has been used to explain the low degree of spontaneous nucleation in the presence of a template~\cite{mohammed2013directing}, but could also be responsible for limiting post-nucleation polymorphism. It will be interesting to explore which features of DNA-tile tubule assembly are responsible for the low polymorphism and reduced spontaneous nucleation, and apply those concepts to the self-assembly of larger subunits, such as DNA-origami colloids.

There are two aspects of assembling seeds that are important to tackle going forward: reducing the complexity of making specific seeds and driving higher assembly yield. Finding more economical ways to construct seeds with larger PU cells should help to not only increase the yield, but also to make it tractable to target larger diameters of tubules. One promising route for this design path is to allow for interactions that are multifarious. One such scheme is algorithmic assembly, which has been shown to provide better complexity scaling of finite-sized assemblies~\cite{rothemund2000program}. It would be interesting to see how this approach can be applied to periodic and self-closing tilings. While it may seem dubious to introduce multifarious interactions due to the presence of off-target structures, recent theoretical studies have shown that multifarious systems of finite-sized assemblies can be driven to high yield~\cite{hubl2025hidden, guttieres2025general} and may even assemble faster and with better fidelity than their fully-addressed counterparts~\cite{hubl2025simultaneous}, though, again, this approach has not been shown for periodic structures. Other ways of driving higher yield may involve controlling the binding energies of the individual interactions or the stoichiometries of the particles~\cite{hubl2025hidden}. This problem of achieving higher yield for subsequent assembly steps would be useful for multistage assemblies, where the yield of initial stages can drastically impact the yield of the final structure~\cite{tikhomirov2017fractal}.

\begin{acknowledgments}
We acknowledge Daichi Hayakawa for starting us down this path. We thank Berith Isaac and Amanda Tiano for their technical support with electron microscopy. TEM images were prepared and imaged at the Brandeis Electron Microscopy facility. This work is supported by the Brandeis University Materials Research Science and Engineering Center (NSF DMR-2011846) and the National Science Foundation (DMR-2214590).

\section*{Data availability}

Data associated with this manuscript, including all TEM images required to generate the state diagram in Figure 2 and the histograms in Figure 3, have been deposited in a permanent Zenodo repository, which will be made available after the manuscript is accepted.

\end{acknowledgments}

\bibliography{main.bib}

\end{document}

\title{Supplemental Appendix for  ``Taming polymorphism of tubule self-assembly using templated growth''}

\author{Sirui Liu}
\affiliation{Martin A. Fisher School of Physics, Brandeis University, Waltham, MA, 02453, USA}

\author{Thomas E. Videb\ae k}
\affiliation{Martin A. Fisher School of Physics, Brandeis University, Waltham, MA, 02453, USA}

\author{W. Benjamin Rogers}
\email{wrogers@brandeis.edu}
\affiliation{Martin A. Fisher School of Physics, Brandeis University, Waltham, MA, 02453, USA}

\maketitle

\setcounter{figure}{0}
\makeatletter 
\renewcommand{\thefigure}{S\arabic{figure}}

% ============ METHODS =====================

\section{Materials and Methods}

\textbf{Folding DNA origami.} To assemble our DNA-origami monomers, we make a solution with 40~nM of p8064 scaffold, 200~nM of each staple strand (Integrated DNA Technologies [IDT]; Nanobase structures 235~\cite{Nanobase} for sequences), and 1x folding buffer. We then anneal this solution using a temperature protocol described below. Our folding buffer, from here on referred to as FoBX, contains 5~mM  Tris Base, 1~mM EDTA, 5~mM NaCl, and X~mM MgCl$_2$. We use a Tetrad (Bio-Rad) thermocycler to anneal our samples.

To find the best folding conditions for each sample, we follow a standard screening procedure to search multiple MgCl$_2$ concentrations and temperature ranges~\cite{wagenbauer2017we, Hayakawa2022}, and select the protocol that optimizes the yield of monomers while limiting the number of aggregates that form. All particles used in this study were folded at 17.5~mM MgCl$_2$ with the following annealing protocol: (i) hold the sample at 65~$^\circ$C for 15 minutes,  (ii) ramp the temperature from 58~$^\circ$C to 51~$^\circ$C with steps of 1~$^\circ$C per hour, (iii) hold at 51~$^\circ$C until the sample can be removed for further processing. \\

\textbf{Agarose gel electrophoresis.} We use agarose gel electrophoresis to assess the folding protocols and purify our samples with gel extraction. We prepare all DNA origami monomer purification gels by bringing a solution of 1.5\%~(w/w) agarose in 0.5x TBE to a boil in a microwave. Once the solution is homogeneous, we cool it to 60~$^\circ$C using a water bath. We then add MgCl$_2$ and SYBR-safe (Invitrogen) to have concentrations of 5.5~mM MgCl$_2$ and 0.5x SYBR-safe. We pour the solution into an Owl B2 gel cast and add gel combs (20 $\upmu$L wells for screening folding conditions or 200~$\upmu$L wells for gel extraction), which cools to room temperature. A buffer solution of 0.5x TBE and 5.5~mM MgCl$_2$, chilled at 4~$^\circ$C for an hour, is poured into the gel box. Agarose gel electrophoresis is run at 110 V for 1.5--2 hours in a 4~$^\circ$C cold room. We scan the gel with a Typhoon FLA 9500 laser scanner (GE Healthcare) at 25~$\upmu$m or 50~$\upmu$m resolution. \\

\textbf{Sample purification.} After folding, we purify our DNA-origami particles to remove all excess staples and misfolded aggregates using gel purification. If the particles have self-complementary interactions, they are diluted 1:1 with 1xFoB2 and held at 45~$^\circ$C for 30 minutes to unbind higher-order assemblies. The folded particles are run through an agarose gel (now at a 1x SYBR-safe concentration for visualization) using a custom gel comb, which can hold around 2~mL of solution per gel. We use a blue fluorescent light table to identify the gel band containing the monomers. The monomer band is then extracted using a razor blade. We place the gel slices into a Freeze ’N Squeeze spin column (Bio-Rad), freeze it in a -20~$^\circ$C freezer for 5 minutes, and then spin the solution down for 5~minutes at 12~krcf. The concentration of the DNA origami particles in the subnatant is measured using a Nanodrop (Thermo Scientific). We assume that the solution consists only of monomers, where each monomer has 8064 base pairs.

Since the concentration of particles obtained after gel purification is typically not high enough for assembly, we concentrate the solution using ultrafiltration~\cite{wagenbauer2017we}. First, a 0.5-mL Amicon 100-kDa ultrafiltration spin column (Millipore) is equilibrated by centrifuging down 0.5~mL of 1xFoB5 buffer at 5~krcf for 7~minutes. Then, the DNA origami solution is added and centrifuged at 14~krcf for 15 minutes. We remove the flow-through and repeat the process until all of the DNA origami solution is filtered. Finally, we flip the filter upside down into a new Amicon tube and spin down the solution at 1~krcf for 2~minutes. The concentration of the final DNA origami solution is then measured using a Nanodrop. \\

\textbf{Multispecies seed assembly} Multispecies seed assembly experiments are conducted with 10~nM and 30~nM DNA-origami particle concentration. For assemblies that are made up of multiple components, the quoted concentration is the total concentration of all subunits, e.g., for a 10-nM experiment with $N$ species, each species has a concentration of 10/$N$~nM. The (1,0) PU cell seed has a 10~nM DNA origami particle concentration. (3,0), (4,0), (6,0), and (7,0) PU cell seed have 30~nM DNA origami particle concentration. Assembly solutions have volumes up to 50~$\upmu$L with the desired DNA origami concentration in a 1xFoB20 buffer. The solution is placed in a 0.2~mL PCR tube and loaded into a thermocycler (Bio-Rad) at 40~$^\circ$C. The thermocycler lid is held at 100~$^\circ$C to prevent condensation of water on the cap of the PCR tube.\\ 

\textbf{Templated tubule assembly} To make templated assembly solutions, we make the solution with pre-annealed seeds and tubule monomers, then gently mix them. After adding the pre-annealed seed into the solution, the MgCl$_2$ concentration is maintained at 1xFoB20 to prevent damage to the assembled structures. The tubule monomer has 10~nM DNA-origami particle concentration. Since we use two species of monomer to assemble a tubule, each species has 5~nM DNA-origami particle concentration. Seeds have concentrations ranging from 0~pm to 8.35~pM. We pipette the assembly solutions into 0.2~mL PCR tubes and assemble them at temperatures ranging from 36~\textdegree C to 39~\textdegree C.\\

\textbf{Labeling tubules with gold nanoparticles.} We first attach thiol-modified ssDNA (5'-HS-C$_6$H$_{12}$-TTTTTAACCA
TTCTCTTCCT-3', IDT) to 10-nm-diameter or 15-nm-diameter gold nanoparticles (AuNP) (Ted Pella). First, we reduce the thiolated strands using tris(2-carboxyethyl) phosphine (TCEP) solution (Sigma-Aldrich) by holding a mixture of 10~mM TCEP (pH 8) and 50~$\upmu$M thiol-DNA at room temperature for one hour on a vortex shaker. We remove excess TCEP with a 10-kDa Amicon filter in four washes of a 1X TE buffer for 30 mins each wash at 4~$^\circ$C; After purification, we store thiolated DNA strands at -20~$^\circ$C until needed. 

To attach thiolated DNA to the AuNP, In a 100~$\upmu$L volume, we mix DNA and gold nanoparticles at a ratio of 500:1 for 10-nm-diameter AuNP and at a 1200:1 ratio for 15-nm-diameter AuNP. We then mix 900~$\upmu$L 1-Butanol into the solution and wait for 20 minutes. We mix 100~$\upmu$L of 1xFoB5 to enhance the salt concentration. We vortex the solution for 30 seconds at the highest intensity, sonicate for 10 minutes, and centrifuge for 20 seconds. To remove excessive thiolated DNA, we wash the solution 5 times with a 0.5~mL Amicon 30kDA ultrafiltration spin column. 

To attach AuNPs to seed, we incorporate handles on the interior edges of the DNA origami subunit with a complementary sequence (5'-AGGAAGAGAATGGTT-3', IDT) to the DNA on the AuNP. After seeded tubules have been assembled, we mix seeded tubule assembly with 8~nM 10~nm AuNP or 4~nM 15~nm AuNP at 1XFoB20 at a 1:1 v/v ratio, then incubate at the assembly temperature for 24~hours. After incubation, samples are ready to be prepared for imaging. \\

\textbf{Negative-stain TEM.} We first prepare a solution of uranyl formate (UFo). We boil doubly distilled water to deoxygenate it and then mix in UFo powder to create a 2\%~(w/w) UFo solution. We cover the solution with aluminum foil to avoid light exposure and vortex it vigorously for 20~minutes, after which we filter the solution with a 0.2~$\upmu$m filter. Lastly, we divide the solution into 0.2~mL aliquots, which are stored in a -80~$^\circ$C freezer until further use.

Before each negative-stain TEM experiment, we take a 0.2~mL UFo aliquot out from the freezer to thaw at room temperature. We add 4~$\upmu$L of 1~M NaOH and vortex the solution vigorously for 15~seconds. The solution is centrifuged at 4~$^\circ$C and 16~krcf for 8~minutes. We extract 170~$\upmu$L of the supernatant for staining and discard the rest.

The TEM samples are prepared using FCF400-Cu grids (Electron Microscopy Sciences). We glow discharge the grid prior to use at -20~mA for 30 seconds at 0.1~mbar, using a Quorum Emitech K100X glow discharger. We place 4~$\upmu$L of the sample on the carbon side of the grid for 1 minute to allow adsorption of the sample to the grid. During this time, 5~$\upmu$L and 18~$\upmu$L droplets of UFo solution are placed on a piece of parafilm. After the adsorption period, the remaining sample solution is blotted on 11~$\upmu$m Whatman filter paper. We then touch the carbon side of the grid to the 5~$\upmu$L drop and blot it away immediately to wash away any buffer solution from the grid. This step is followed by picking up the 18~$\upmu$L UFo drop onto the carbon side of the grid and letting it rest for 30 seconds to deposit the stain. The UFo solution is then blotted and any excess fluid is vacuumed away. Grids are allowed to dry for a minimum of 15 minutes before insertion into the TEM. 

We image the grids using an FEI Morgagni TEM operated at 80 kV with a Nanosprint5 CMOS camera (AMT). Micrographs are acquired between x1,100 to x14,000 magnification. 

To prevent the overlap of structures, we dilute the assembly as needed. For example, we dilute the tubule assembly to 5~nM, and dilute the seed assembly to 10~nM before making the TEM grids.\\

\textbf{Assemble UV-weldable seed} UV-weldable seeds are assembled at 40~nM total DNA origami particle concentration. Monomers are mixed in a stoichiometric ratio based on how they appear in the seed. The seed was assembled at 15~mM MgCl$_2$ and 50~mM NaCl with the following protocol: (i) ramp the temperature from 50~\textdegree C to 42~\textdegree C with steps of 1~\textdegree C per hour, (ii) hold at 42~\textdegree C for 24 hours. Assembly volumes are 50~$\upmu$L placed in 0.2~mL PCR tubes.   \\

\textbf{UV welding and seed purification} After finishing the assembly protocol, we shine the 310~nm UVC light (Boston Electronics) on the assembly solution for 20~minutes. We prepare a gel by bringing a solution of 0.5\% (w/w) agarose in 0.5x TBE to a boil in a microwave. Once the solution is homogeneous, we cool it to 60~\textdegree C using a water bath. Then, we add MgCl$_2$, NaCl, SYBR-safe (Invitrogen) to have concentrations of 10~mM MgCl$_2$, 50mM NaCl, and 0.5x SYBR-safe. We pour the solution into an Owl B2 gel cast and add a customized gel comb (each well is 1~mm thick and can hold 80~$\upmu$L solutions), which is cooled to room temperature. A buffer solution of 0.5x TBE, 10~mM MgCl$_2$, and 50~mM NaCl, chilled at 4~\textdegree C for an hour, is poured into the gel box. Gel electrophoresis is run at 70~V for 7~hours in a 4~\textdegree C cold room. We change gel buffers every 90~min.

After finishing gel electrophoresis, we scan the gel with the Typhoon FLA 9500 laser scanner to find the position of each band and cut out the desired seed type with a razor blade. We place the gel slices into a Freeze 'N Squeeze spin column (Bio-Rad), freeze it in a -20~\textdegree C freezer for 60~minutes, then thaw for 10~minutes, and spin the solution down for 5~minutes at 10~krcf. The solution is concentrated using ultrafiltration as before, expect we equlibrate the spin column with a buffer of 1xFoB10 with and added 50~mM NaCl.

% ============ END oF METHODS =====================

\section{UV welding}

\begin{figure*}[!bh]
    \centering
    \includegraphics[width=\linewidth]{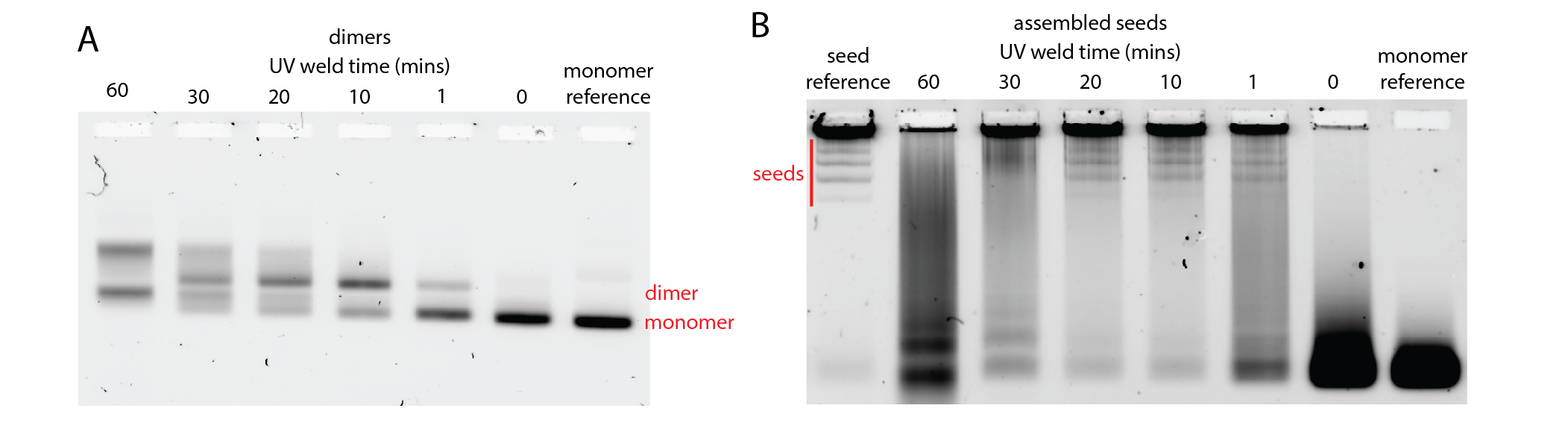}
    \caption{\textbf{UV welding stabilizes assembled structures.} (A) Testing the melting of UV-welded dimers using gel electrophoresis. Dimer assemblies are irradiated for varying amounts of time and compared to a monomer reference. (B) Testing the melting of seed assemblies using gel electrophoresis. Seed assemblies are irradiated for varying amounts of time and compared to both a monomer reference and an assembled seed reference.}
    \label{SFig:uv_proof_of_concenpt} 
\end{figure*}

We first conduct a dimerization test to determine if UV welding can stabilize bound subunits. We make two monomer types, which are passive on sides 1 and 2, and have the A and A* interaction from Table~\ref{tab:6bp_side_interactions} on side 3. We prepare assemblies with 20~nM monomers and 20~mM MgCl$_2$ and incubate at room temperature for 45 minutes to promote dimer formation. We shine UVC light on the assembly for various durations from 0 minutes to 60 minutes. To test the stability of the UV-welded bond, we dilute each sample to 5~nM monomer and 5~mM MgCl$_2$ and heat to 50~\textdegree C for 45 minutes. We then run the samples in a gel to examine the assembly result (Fig.~\ref{SFig:uv_proof_of_concenpt}A). We notice that there is a monomer to dimer transition as we increase the UV light exposure time to the assembly from 0 minutes to 10 minutes. After that, the monomer and dimer bands shift toward the loading pocket. This may indicate damage to the DNA structure~\cite{gerling2018sequence}. 

Next, we test if UV welding can stabilize an assembled seed. We prepare UV-weldable seeds in assembly solutions of 30~nM monomers and 15~mM MgCl$_2$. We shine the UVC light on the assemblies with various durations from 0 minutes to 60 minutes. Then, we dilute each sample to 10~nM DNA origami particle concentration and 5~mM MgCl$_2$ concentration, and we heat the sample to 42~\textdegree C for 60 minutes. We then run the samples in a gel to examine the assembly result (Fig.~\ref{SFig:uv_proof_of_concenpt}B). The first lane from the left is a UV-weldable seed reference. We see results similar to the dimer case above. At 0 minutes, the assemblies dissociate into monomers. For 1 to 20 minutes of UVC exposure, the seed assemblies remain intact. For longer exposure, we see the assembly bands change into smears, which may indicate excessive damage to the assemblies. From these results, we decided to UV-weld assemblies for 20 minutes to create stable seeds.

%UV seed melting experiment

\begin{figure*}[!th]
    \centering
    \includegraphics[width=\linewidth]{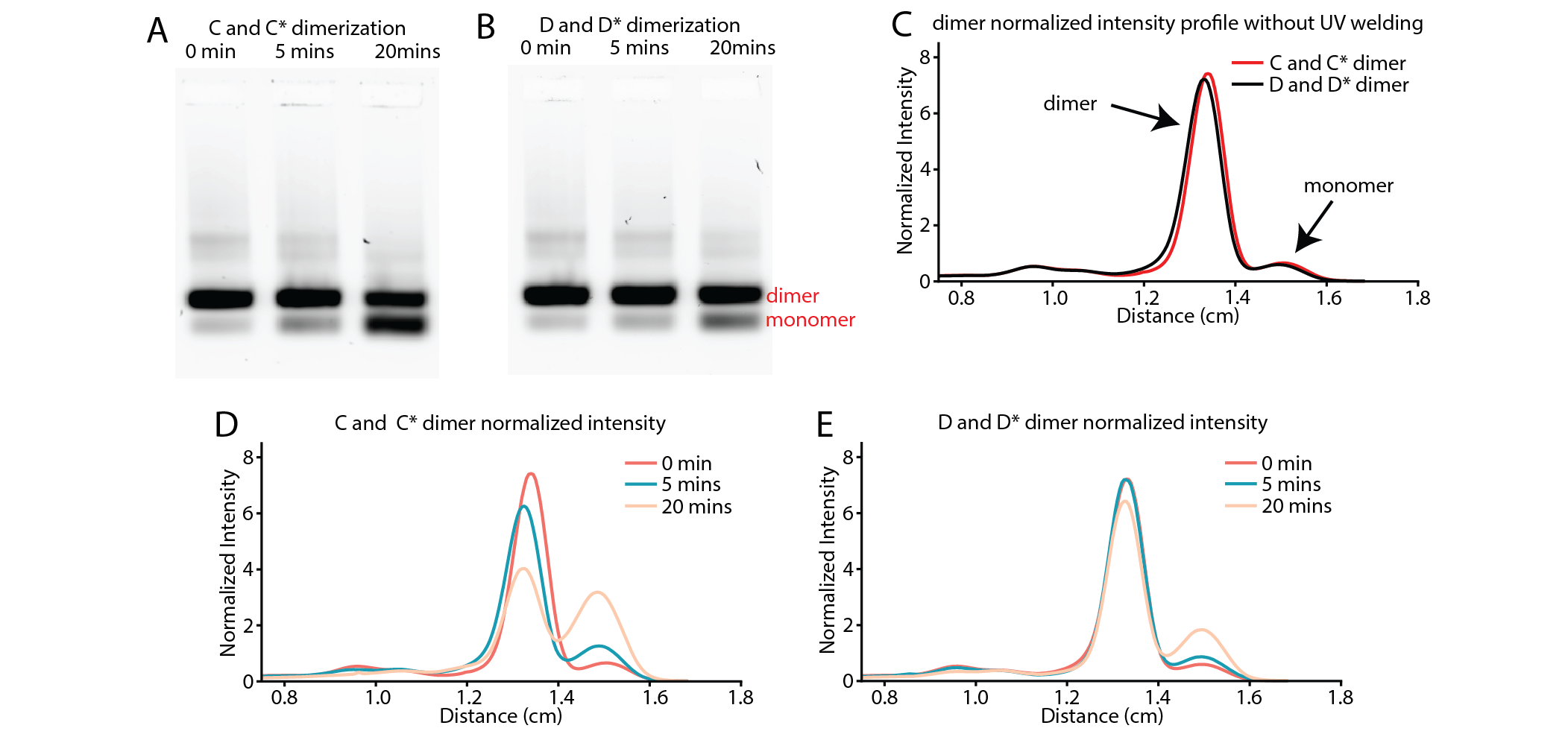}
    \caption{\textbf{UV light damages interaction sequence.} Gel electrophoresis results of C/C* (A) and D/D* (B) dimerization. (C) Normalized intensity profile of C/C* and D/D* dimers without UVC exposure. (D) Comparison of normalized C/C* intensity profiles when monomer C* is exposed to UVC light for 0, 5, and 20 minutes. (E) Comparison of normalized D/D* intensity profiles when monomer D* is exposed to UVC light for 0, 5, and 20 minutes.
 }
    \label{SFig:uv_dimer_damage}
\end{figure*}

When trying to grow tubules off of the UV-welded seed, we observe that tubule monomers do not attach to the seed. This observation suggests that UV light can reduce the binding strength of our ssDNA interactions. Upon inspection, the sequences used on the edge of the seed contained a few adjacent thymines or cytosines. To test if these thymine and cytosine pairs were to blame, we perform a dimerization experiment using different interaction sequences on side 3. In this experiment, all monomer species are passivated on sides 1 and 2, while side 3 contains the interactions C/C* or D/D* from Table~\ref {tab:6bp_side_interactions}. We prepare monomers in 20~mM MgCl$_2$ and then shine UV light on the monomers containing interactions C and D* for 0, 5, or 20 minutes. Each handle in C contains adjacent thymines or cytosines, whereas only one of six handles in D* contains adjacent cytosines. After UV exposure, we mix each monomer with its complementary monomer and incubate the mixture at room temperature for 20 hours. We assess the results using gel electrophoresis, as shown in Fig.~\ref{SFig:uv_dimer_damage}A and B. While both dimer pairs seem to behave the same after 5 minutes of UVC exposure, at 20 minutes, the C/C* interaction has a lower binding affinity, shown by the increased intensity of the monomer band (Fig.~\ref{SFig:uv_dimer_damage}D,E).

% UV dimer sequence

%\clearpage

\section{Determining tubule type}

\begin{figure*}[!th]
    \centering
    \includegraphics[width=\linewidth]{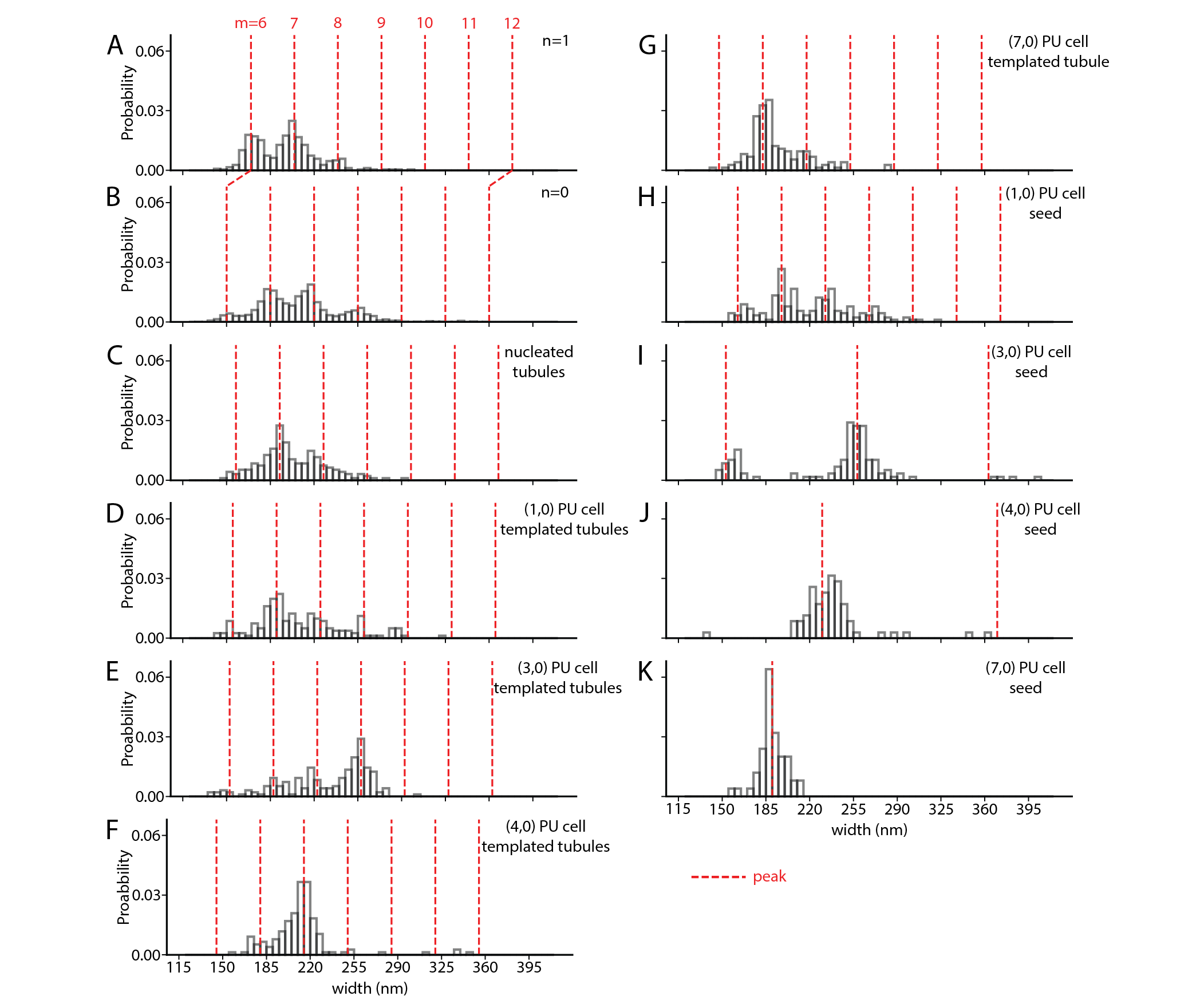}
    \caption{\textbf{TEM multispecies tubule analysis}. (A) and (B) are the width distributions of tubules with $n$=1 and $n$=0. Red lines show predicted tubules for various $m$ numbers. (A) has 471 tubule segments. (B) has 1991 tubule segments. (C)-(K) are the width distribution for various $n$=0 samples. 
    }
    \label{SFig:tubule_type_binning}
\end{figure*}

Every tubule and seed can be identified with a unique pair of numbers, ($m,n$), as described in the main text. To identify these numbers, we follow the method used in Videb{\ae}k et al.~\cite{videbaek2024economical}. The $n$ number of tubules can be identified by the pattern of the triangular lattice and its pitch relative to the tubule axis. The $m$ number can be a bit less straightforward to obtain since it is related to the circumference, and when the tubule becomes deposited on the grid, it has some uncertainty associated with it based on the sample preparation before TEM imaging. To get around this, we use the fact that when looking at the distributions of the widths of tubules measured in TEM, there are well-defined peaks for a given $n$ number (Fig.~\ref{SFig:tubule_type_binning}A,B). We then use the theoretical circumference of a tubule $L\sqrt{m^2 + n^2 + mn}$, where $L$ is the edge length of a monomer. The width of the tubule on the TEM grid is, if the tubule has no thickness when deposited, half of this circumference. Of course, there is some thickness, so we take an expression for the tubule width on the grid to be $w = L\sqrt{m^2 + n^2 +mn}/2 + b$, where $b$ is used to account for some thickness of the structure. We use the peaks of the width distribution in Fig.~\ref{SFig:tubule_type_binning}B to fit the monomer edge length and the thickness parameter, getting $L=70$ nm and $b=-60$ nm. This allows us to make predictions for tubule widths for structures that do not have well-defined peaks (the red vertical lines in Fig.~\ref{SFig:tubule_type_binning}). With these predictions, we then identify the $m$ number of a tubule by seeing which predicted tubule width the measured width is closest to.

For subsequent experiments, the staining of the TEM grids may be slightly different, so we expect that our value of $b$ may change slightly. For all of the $n=0$ tubules for these experiments, we always have enough statistics to discern a single peak that we can correspond to a tubule type. Using that peak, we remeasure $b$ and make new predictions for the tubule widths (Fig.~\ref{SFig:tubule_type_binning}C-G). For identifying the multispecies seeds, we make one other change, which is that we only predict tubule states that have ($m,n$) values that are allowed by our PU cells (Fig.~\ref{SFig:tubule_type_binning}H-K).

\clearpage

\begin{figure*}[!th]
    \centering
    \includegraphics[width=\linewidth]{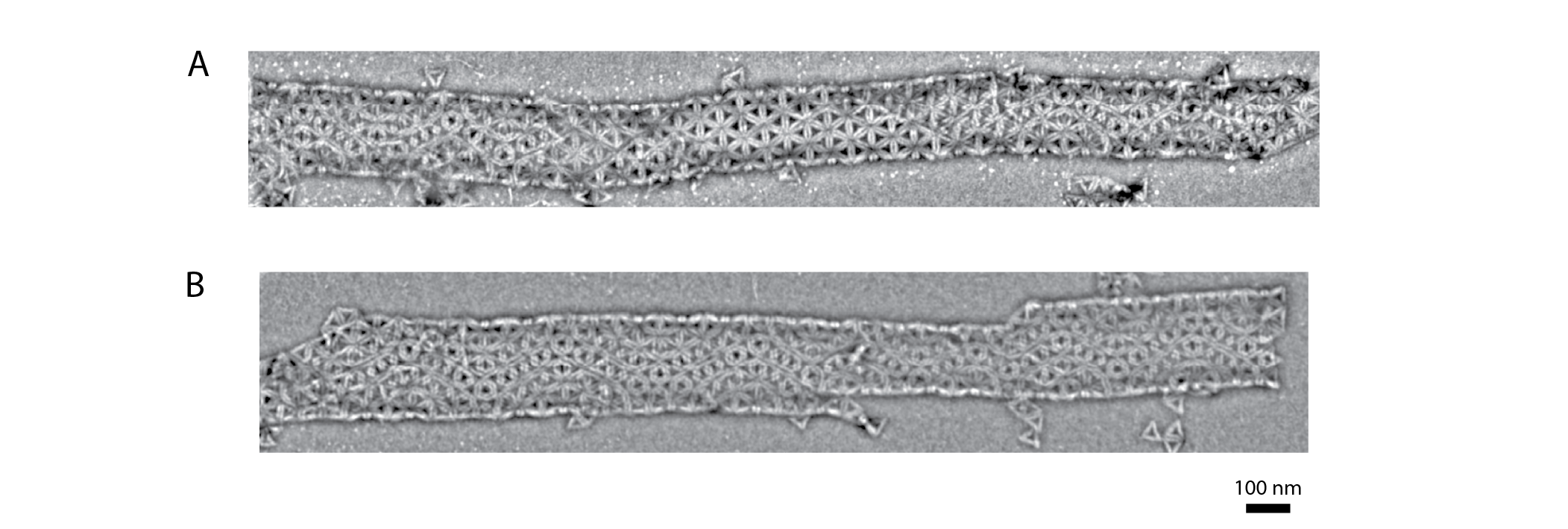}
    \caption{\textbf{Examples of nucleated tubules with changed types}. }
    \label{SFig:unseeded change type}
\end{figure*}

\begin{figure*}[!th]
    \centering
    \includegraphics[width=\linewidth]{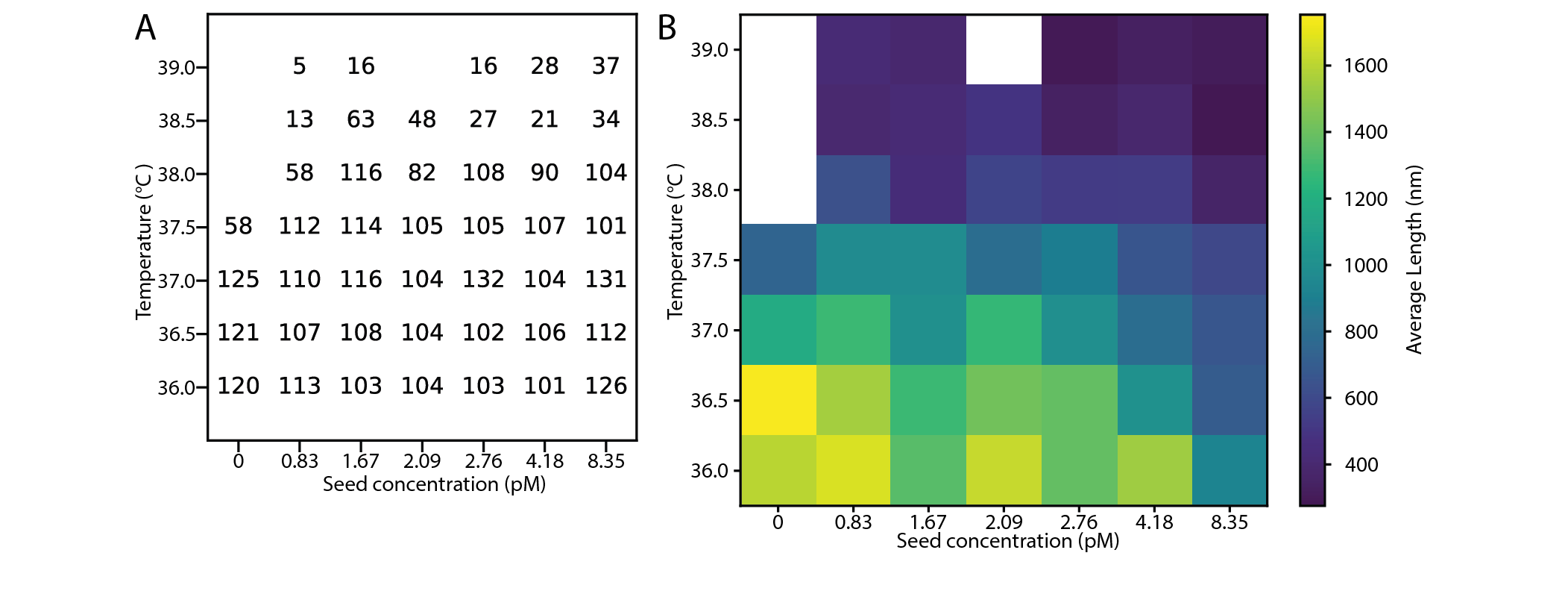}
    \caption{\textbf{Assembly count and tubule length for templated growth experiments}. (A) Plot of the number of assemblies measured for each experimental condition. (B) The color map shows the average tubule length of seeded assemblies with various concentrations of (1,0) PU cells and incubated at different temperatures.}
    \label{SFig:average_length}
\end{figure*}

\begin{figure*}[!th]      
    \centering
    \includegraphics[width=\linewidth]{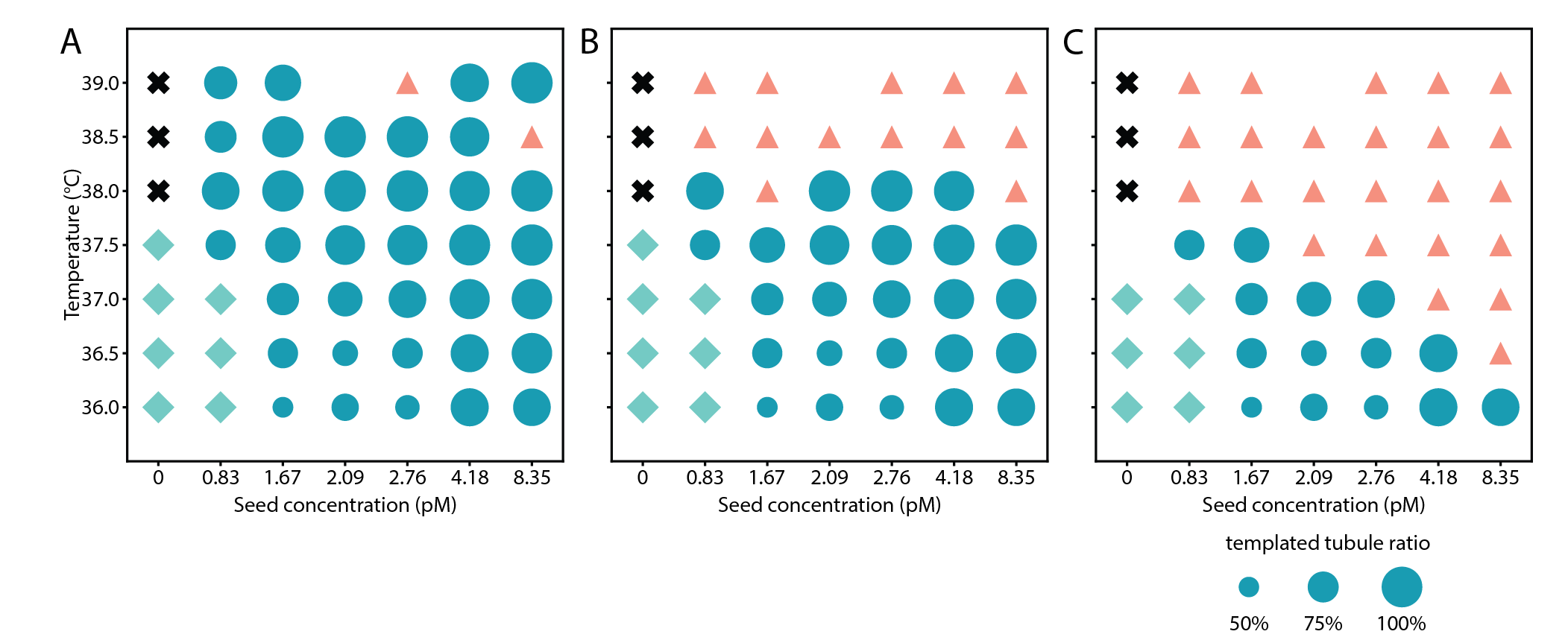}
    \caption{\textbf{State diagram with different cutoff length}. The state diagrams show the same states as in the main text but with changing cutoff lengths separating the templated tubule state and the short-templated tubule state: average tubule length cutoffs are 300~nm (A), 500~nm (B), and 900~nm (C). The assembly with 0~pM seed at 37.5~$^\circ$C has an average tubule length of 739~nm, so when the cutoff is 900~nm, this assembly does not belong to any state.}
    \label{SFig:phase_diagram_different_cutoff}
\end{figure*}

\begin{figure*}[!th]      
    \centering
    \includegraphics[width=\linewidth]{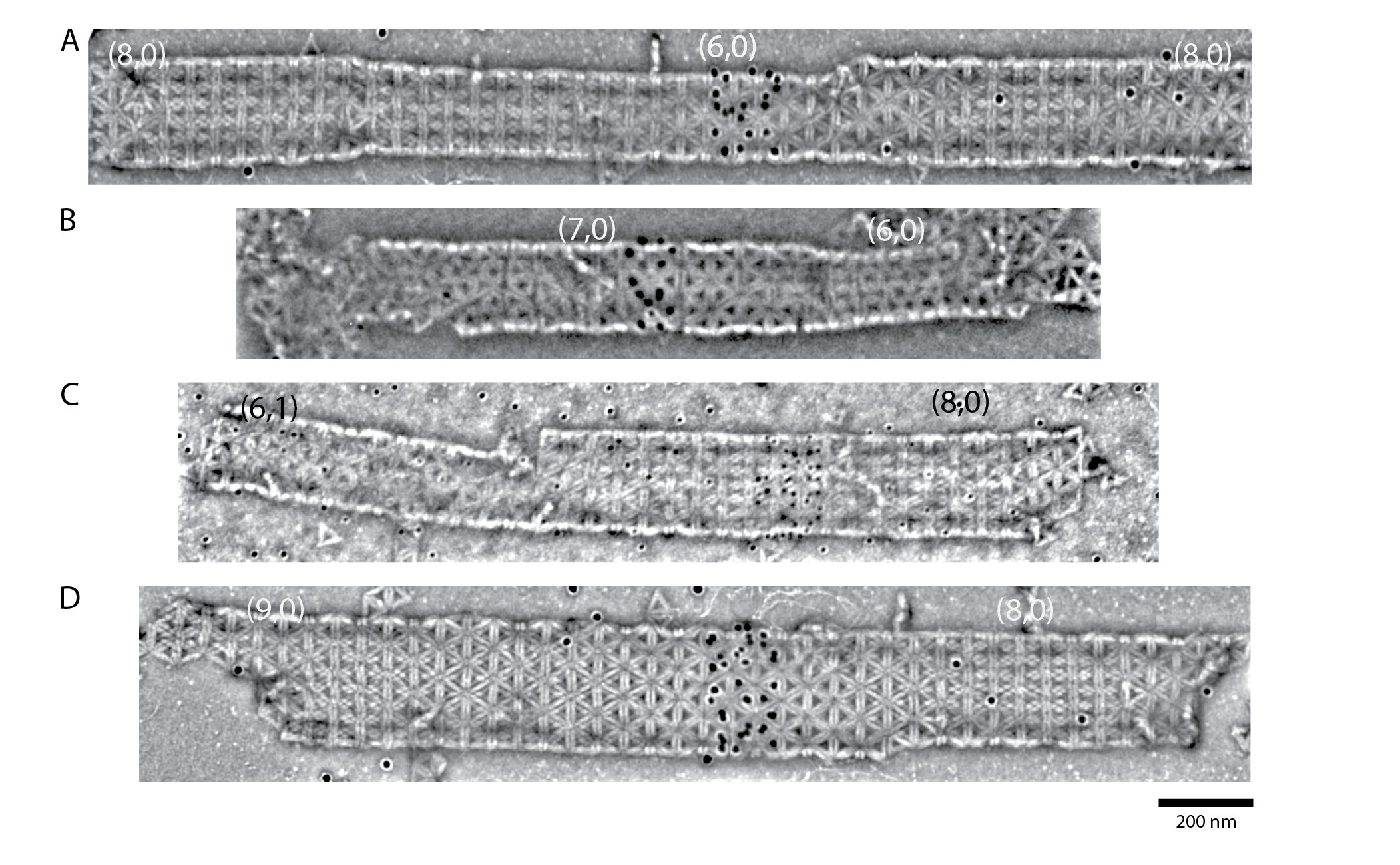}
    \caption{\textbf{Examples of templated tubules with changed tubule types.} All images are label with their ($m$,$n$) numbers and share the same scale bar. (A) A (6,0) templated tubule grown from a (3,0) PU cell seed. (B) A (7,0) templated tubule grown from a (7,0) PU cell seed. (C) A (8,0) templated tubule grown from a (4,0) PU cell seed. (D) A (9,0) templated tubule grown from a (3,0) PU cell seed. }
    \label{SFig:seeded_tubule_type_change}
\end{figure*}

\begin{figure*}[!th]      
    \centering
    \includegraphics[width=\linewidth]{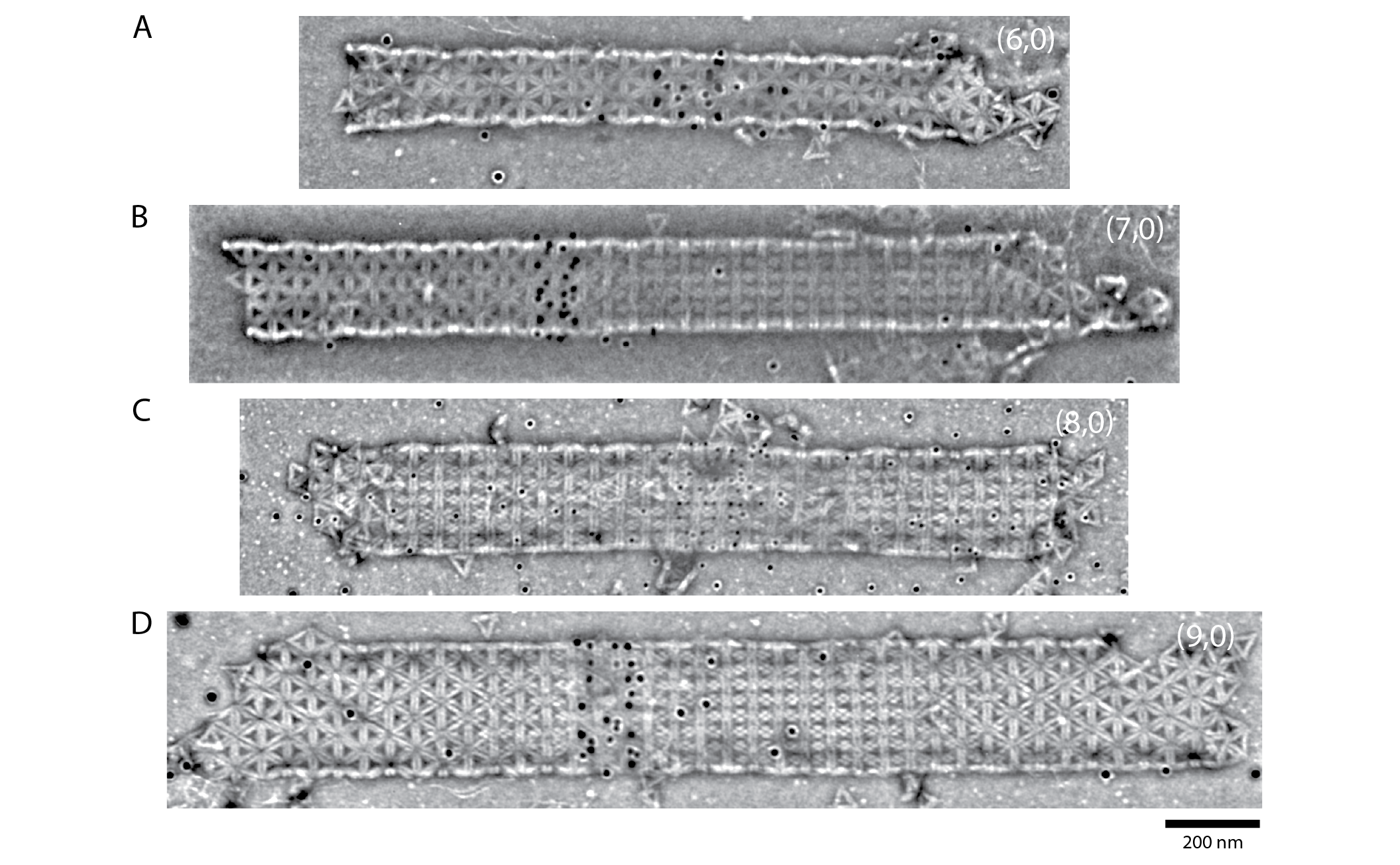}
    \caption{\textbf{Examples of templated tubules with an unchanged tubule type.} All images are labeled with their ($m$,$n$) numbers and share the same scale bar. (A) A (6,0) templated tubule grown from a (3,0) PU cell seed. (B) A (7,0) templated tubule grown from a (7,0) PU cell seed. (C) A (8,0) templated tubule grown from the (4,0) PU cell seed. (D) A (9,0) templated tubule grown from a (3,0) PU cell seed.}
    \label{SFig:seeded_tubule_type_remain}
\end{figure*}

\begin{figure*}[!th]
    \centering
    \includegraphics[width=\linewidth]{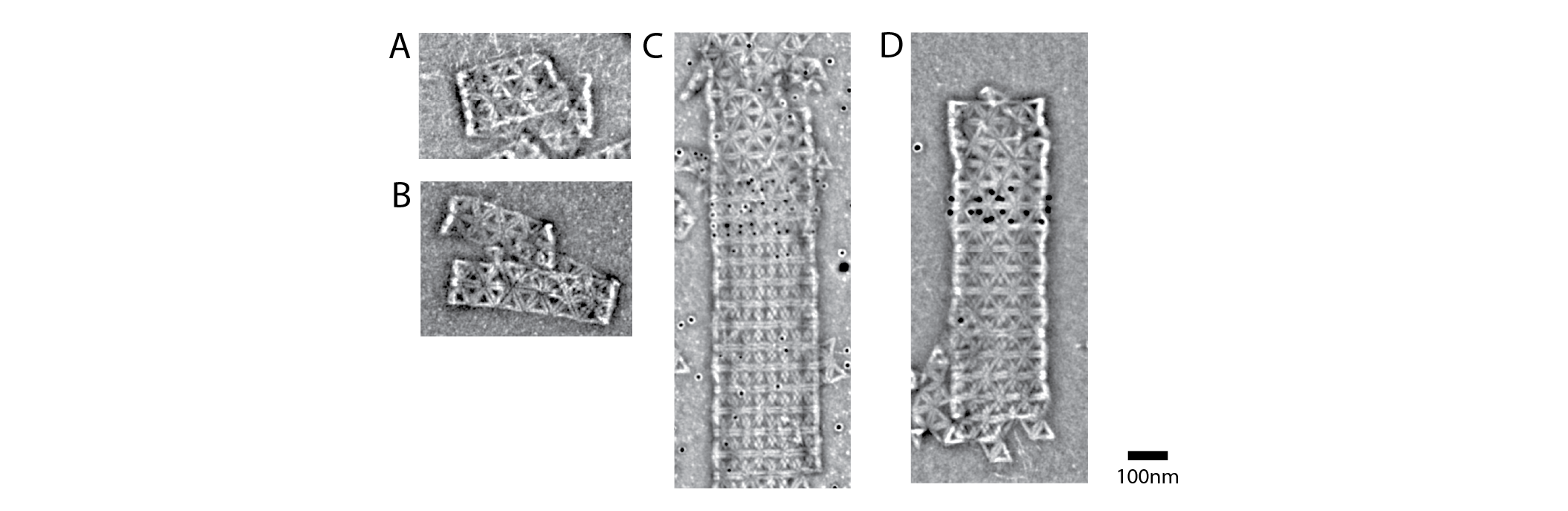}
    \caption{\textbf{Unclosed seed assemblies can cause off-target tubule states}. For high complexity seeds, assembly can miss a closure point and will remain as an unclosed sheet. (A) An example of a (4,0) PU cell assembly forming an unclosed seed. (B) An example of a (7,0) PU cell assembly forming an unclosed seed. We  (C) A (9,0) templated tubule from the (4,0) PU cell seed assembly. (D) A (8,0) templated tubule from the (7,0) PU cell seed assembly. All images share the same scale bar. All images have been bandpass filtered.}
    \label{SFig:high_specificity}
\end{figure*}

% Unseeded, 36c, nucleated tubules
\begin{figure*}[!th]
    \centering
    \includegraphics[width=\linewidth]{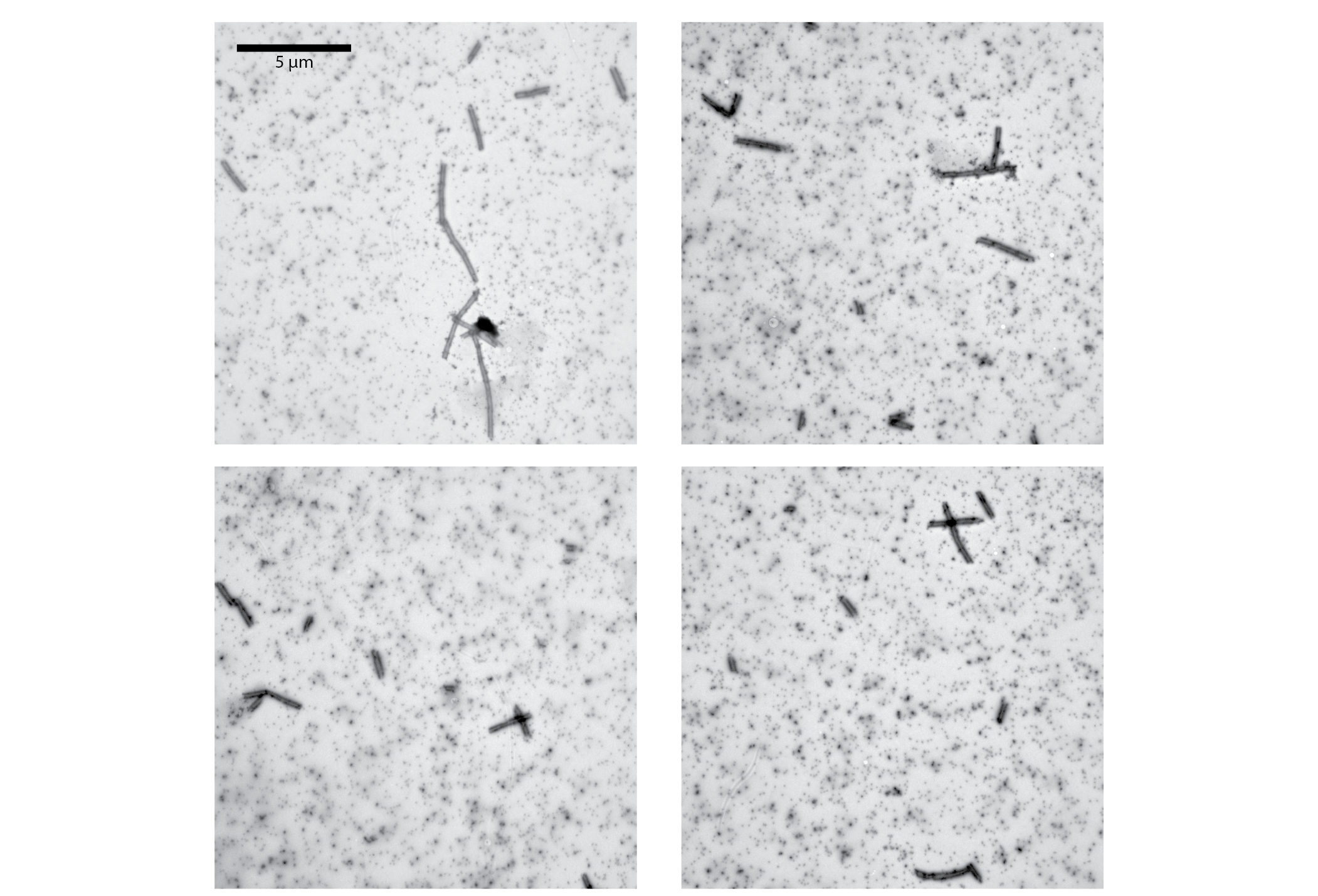}
    \caption{\textbf{Overview of unseeded tubule assembly at 36~\textdegree C.} The TEM images primarily show tubules, unbound monomers, and oligomers. The assembly is in the nucleated tubules state. All images share the same scale bar.}
    \label{SFig:unseeded_36c}
\end{figure*}

% Unseeded 38c, 'no assembly' phase
\begin{figure*}[!th]
    \centering
    \includegraphics[width=\linewidth]{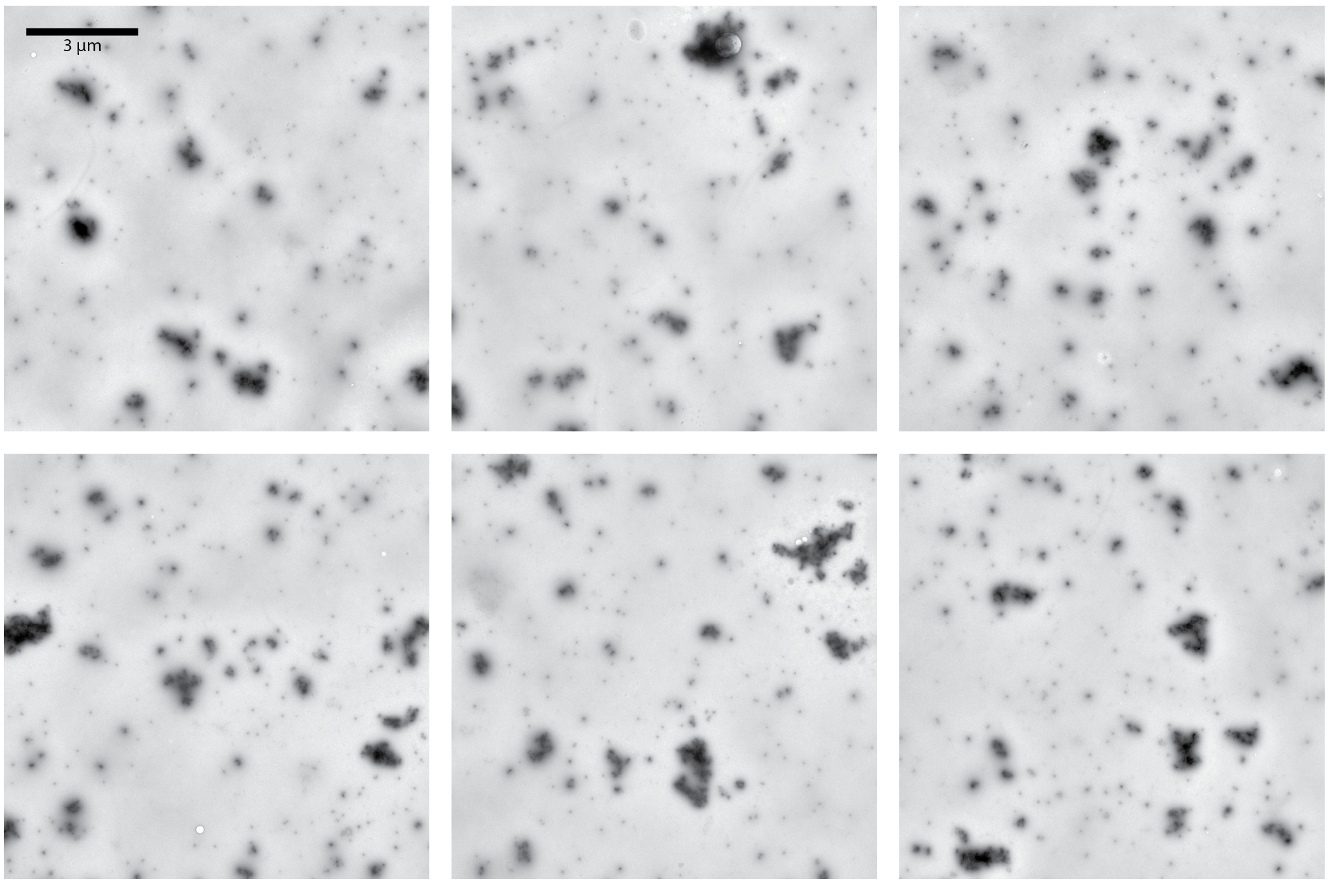}
    \caption{\textbf{Overview of the untemplated tubule assembly at 38~\textdegree C.} There is no tubule at this assembly condition. The TEM shows monomers, oligomers, and aggregates. The assembly is in the no assembly state. All images share the scale bar. }
    \label{SFig:unseeded_38c}
\end{figure*}

% seeded 0.1nM 36c unseeded phase
\begin{figure*}[!th]
    \centering
    \includegraphics[width=\linewidth]{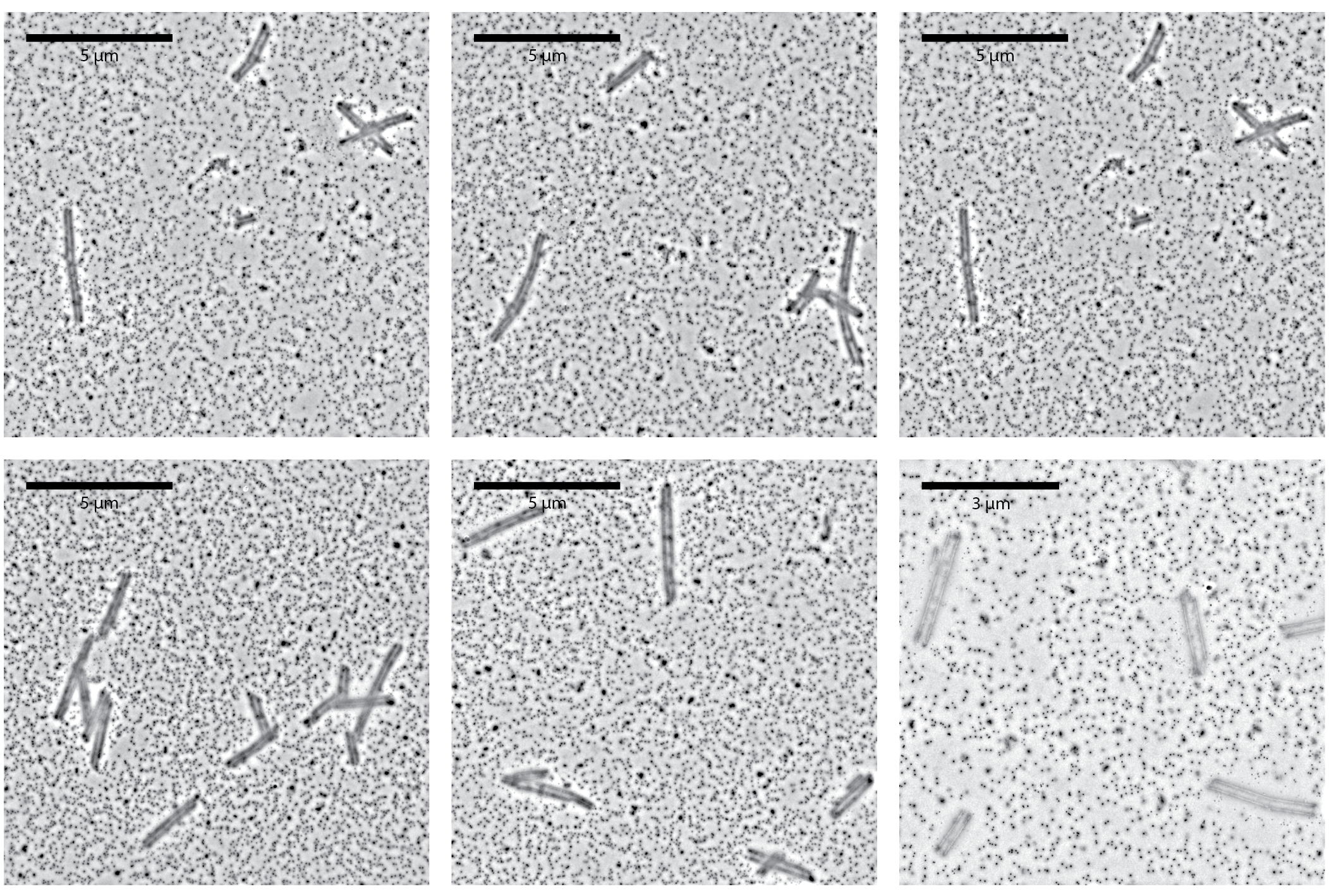}
    \caption{\textbf{Overview of seeded tubule assembly with 0.83~pM seed at 36~\textdegree C.} The TEM images show templated tubules, nucleated tubules, and unbound monomers. The majority of assemblies are nucleated tubules. The black dots in the background are gold nanoparticles (AuNP). Images have been bandpass filtered. }
    \label{SFig:seeded_0.1nM_36c}
\end{figure*}

% seeded 0.1nM, 37.5c seeded phase
\begin{figure*}[!th]
    \centering
    \includegraphics[width=\linewidth]{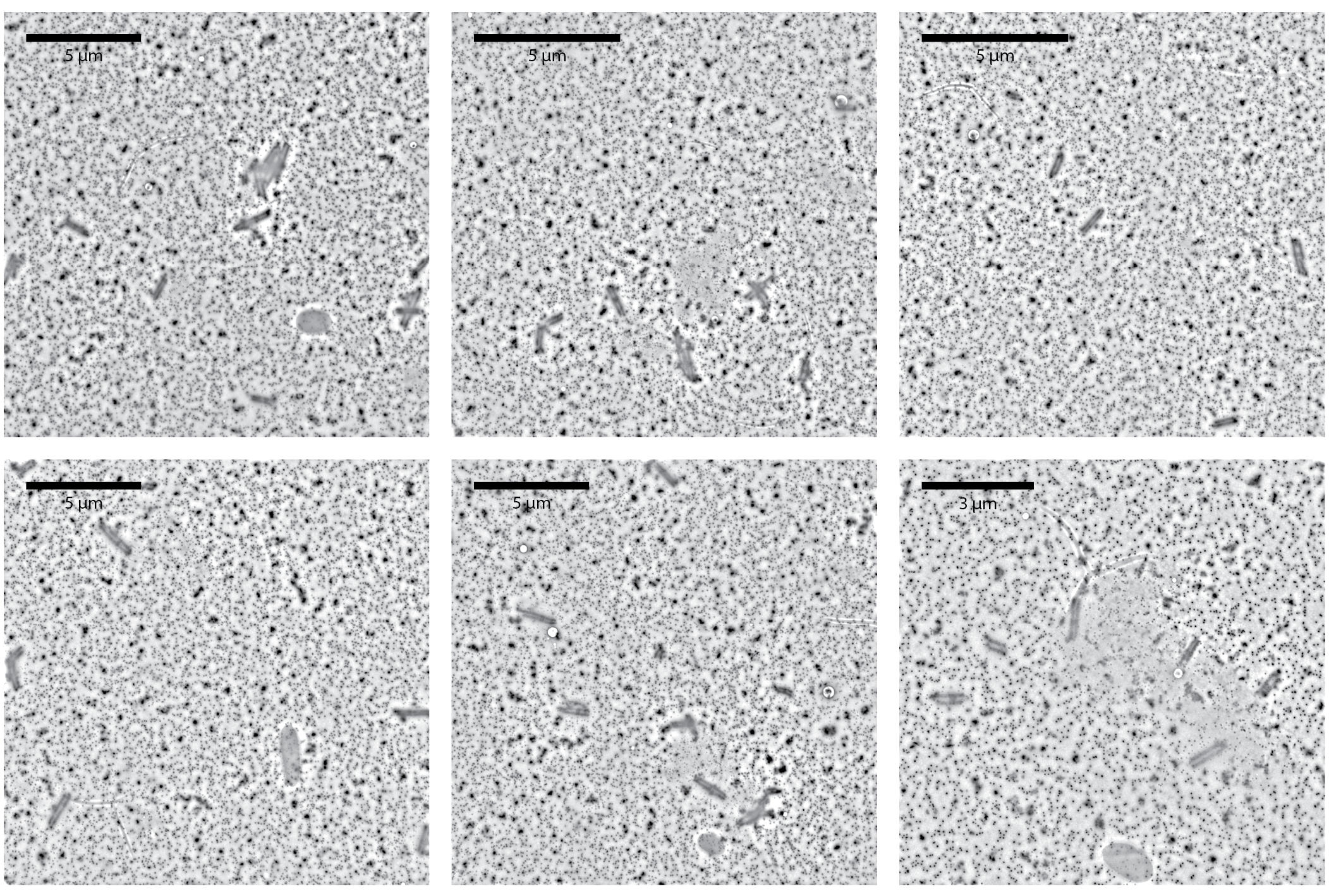}
    \caption{\textbf{Overview of templated tubule assembly with 0.83~pM seed at 37.5~\textdegree C.} The TEM images show templated tubules, nucleated tubules, and unbound monomers. The majority of assemblies are templated tubules.  The black dots in the background are gold nanoparticles (AuNP). These images have been bandpass filtered. }
    \label{SFig:seeded_0.1nM_37.5c}
\end{figure*}

% seeded 0.5nM 39c short tubule phase
\begin{figure*}[!th]
    \centering
    \includegraphics[width=\linewidth]{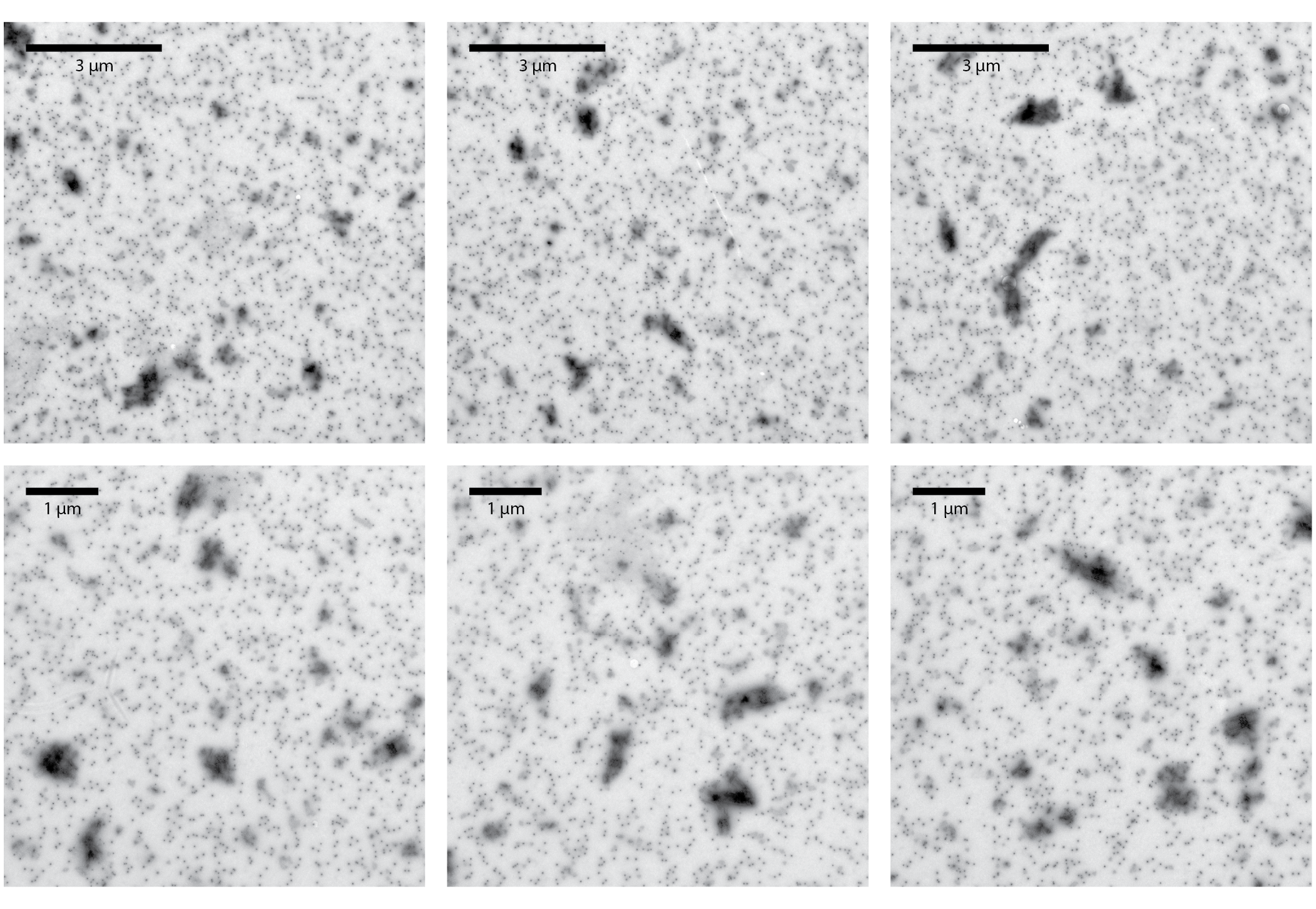}
    \caption{\textbf{Overview of seeded tubule assembly with 4.18~pM seed at 39~\textdegree C.} The TEM images primarily show templated tubules, unbound monomers, and oligomers. Spontaneous nucleation cannot happen at this temperature. The assembly is in the short-templated tubules state. The black dots in the background are gold nanoparticles (AuNP). }
    \label{SFig:seeded_1nM_39c}
\end{figure*}

% seeded 1nM 36c, seeded tubule phase
\begin{figure*}[!th]
    \centering
    \includegraphics[width=\linewidth]{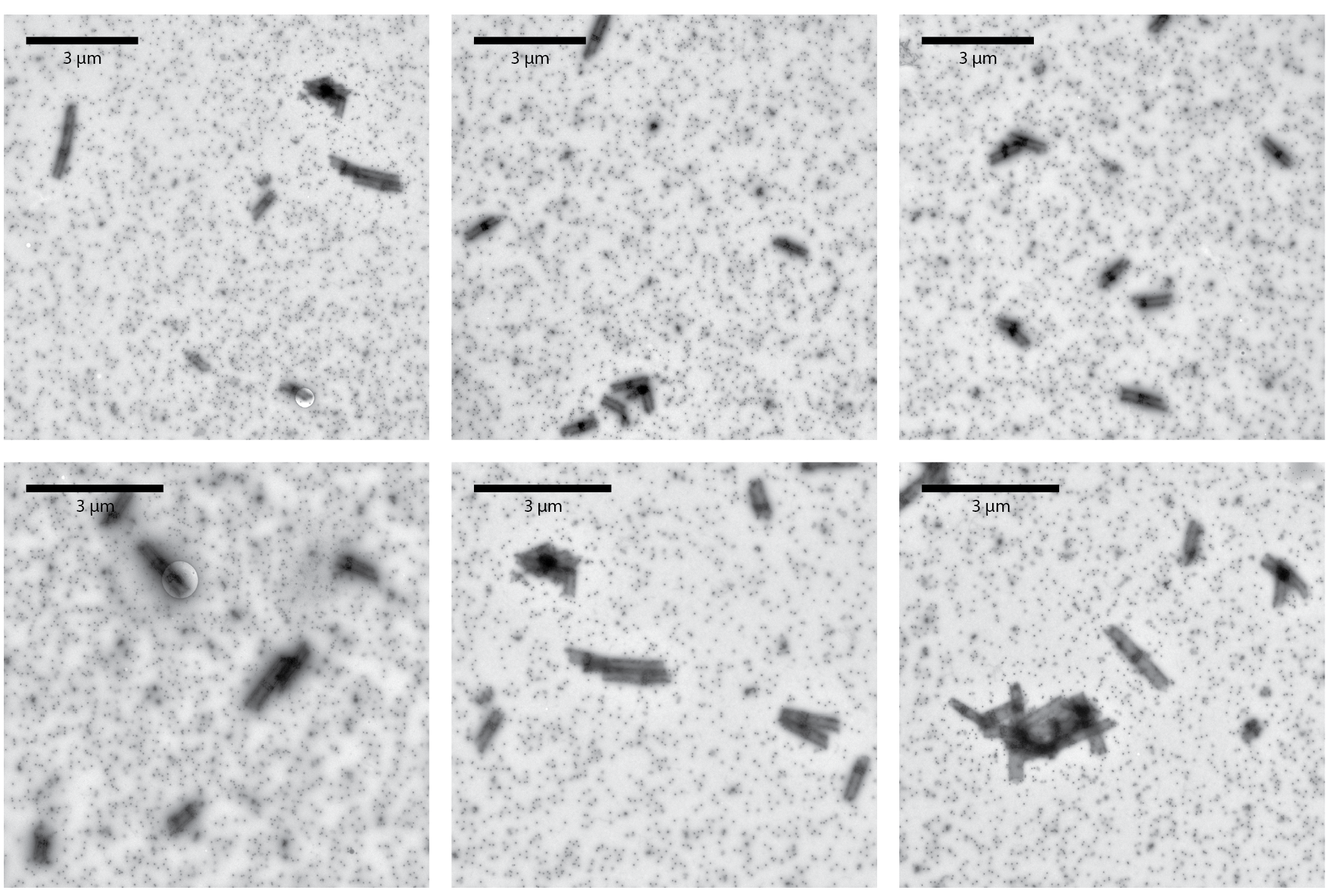}
    \caption{\textbf{Overview of templated tubule assembly with 8.35~pM seed at 36~\textdegree C.} The TEM images primarily show templated tubule, unbound monomers, and oligomers. The assembly is in the templated tubule state.  The black dots in the background are gold nanoparticles (AuNP). }
    \label{SFig:seeded_1nM_36c}
\end{figure*}

% seeded 0.33nM 37c, seeded tubule phase
\begin{figure*}[!th]
    \centering
    \includegraphics[width=\linewidth]{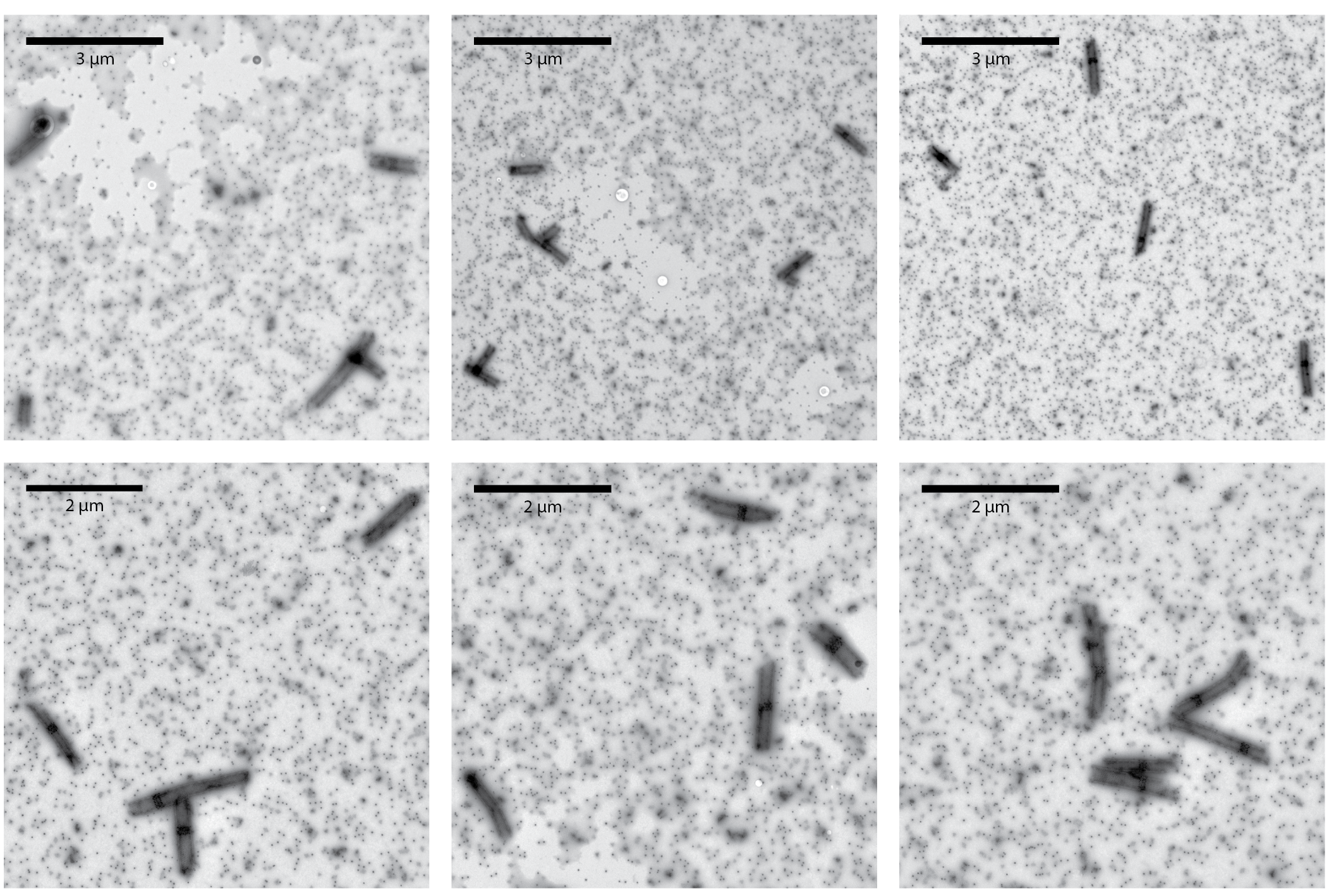}
    \caption{\textbf{Overview of templated tubule assembly with 2.76~pM seed at 37~\textdegree C.} The TEM images primarily show templated tubules, nucleated tubules, unbound monomers, and oligomers. The majority of tubules have a seed. The assembly is in the templated tubules state.  The black dots in the background are gold nanoparticles (AuNP). }
    \label{SFig:seeded_0.33nM_37c}
\end{figure*}

% 3 Layers tubulette gel

\begin{figure*}[!th]
    \centering
    \includegraphics[width=\linewidth]{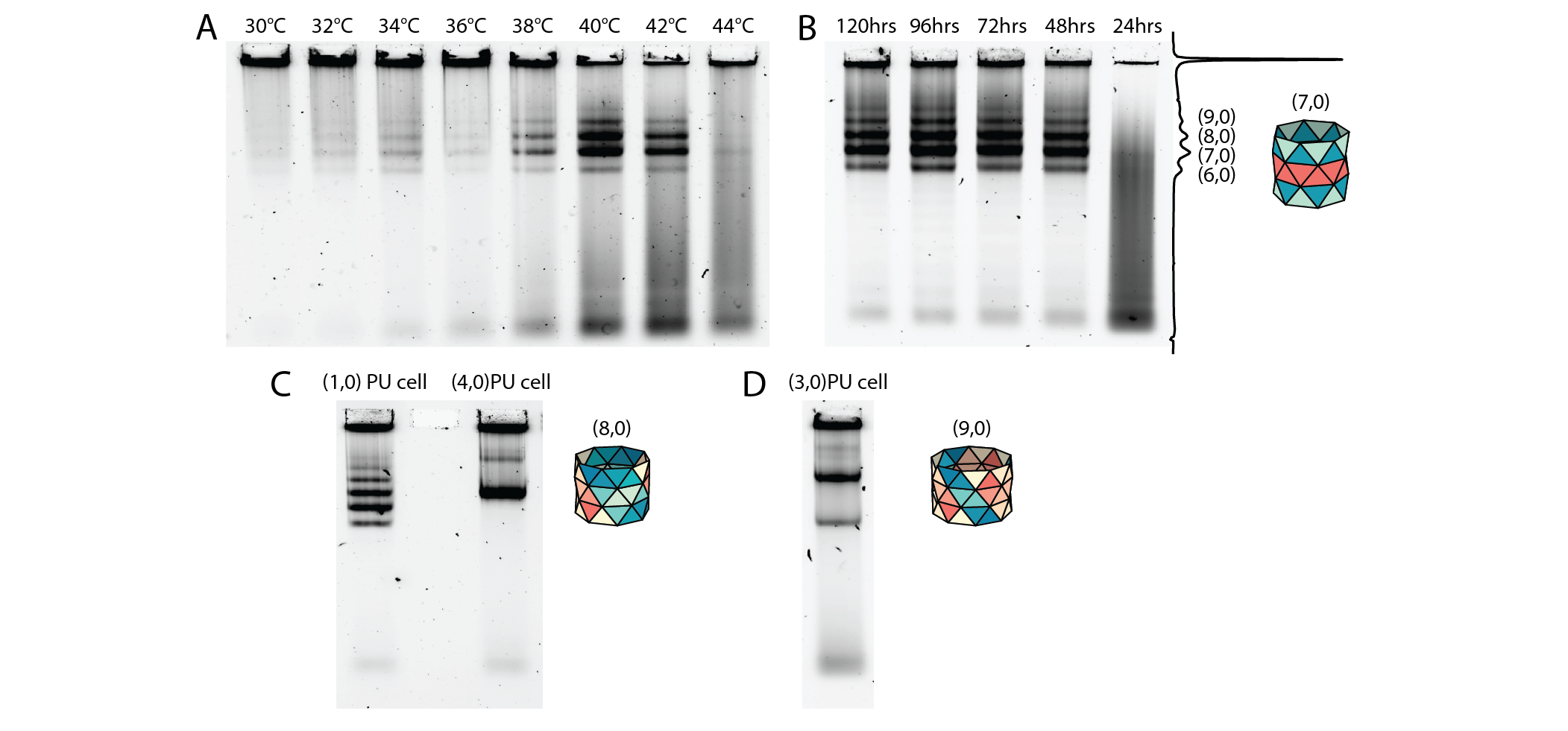}
    \caption{\textbf{Gel electrophoresis of three-layer seeds.} (A) Gel of (1,0) PU cell seed assemblies incubated at varying temperatures. DNA-origami particle concentration is 10~nM. We incubate for 96~hrs. (B) Gel of (1,0) PU cell seed assemblies for varying assembly times. DNA-origami particle concentration is 10~nM. We incubate at 40~\textdegree C. The intensity profile on the right comes from the ``96 hrs'' lane of the gel. (C) Gel electrophoresis of a (4,0) PU cell seed assembly. We incubate 30~nM DNA at 40~\textdegree C for 7 days, and dilute to 10~nM before loading the gel. An assembly of the (1,0) PU cell seed is run as a comparison. (D) Gel electrophoresis of a (3,0) PU cell seed assembly. We incubate 30~nM DNA at 40~°C for 4 days and dilute to 10~nM before loading the gel. All gels have 0.5~\% agarose and 20~mM MgCl$_2$}
    \label{SFig:3layer_seed_gel}
\end{figure*}

% 2 Layers tubulette gel
\begin{figure*}[!th]
    \centering
    \includegraphics[width=\linewidth]{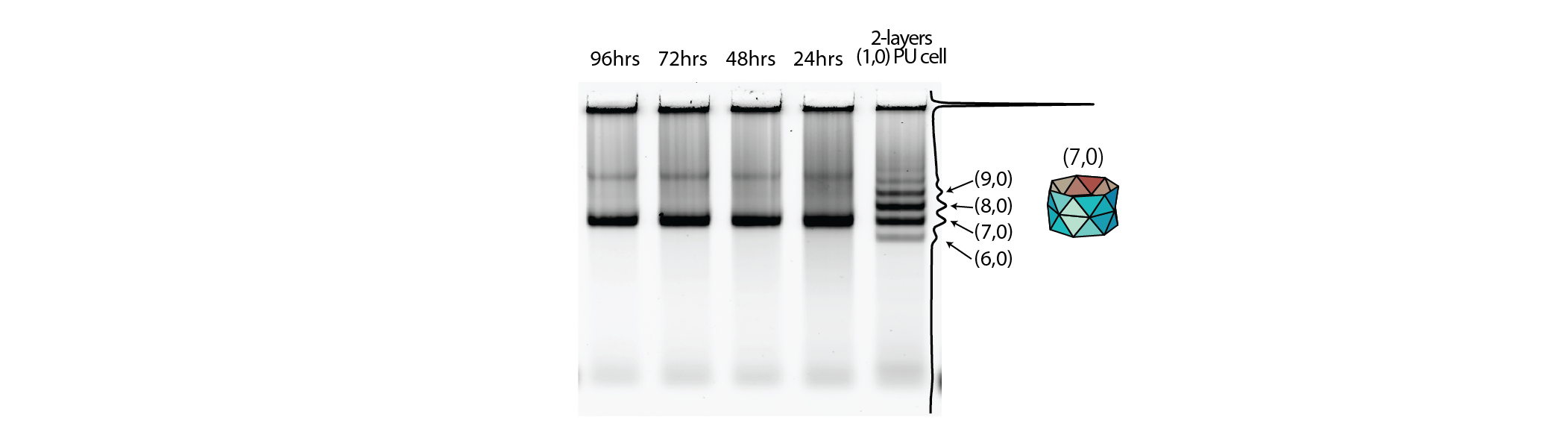}
    \caption{\textbf{Gel electrophoresis of the two-layer seed.}  Gel of (7,0) PU cell seeds for varying assembly time. We assemble the (7,0) PU cell seed with 30~nM monomers, then dilute to 10~nM monomers before loading the gel. The right-most lane is a two-layer (1,0) PU cell seed for reference.  All gels have 0.5~\% agarose and 20~mM MgCl$_2$}
    \label{SFig:2layer_seed_gel}
\end{figure*}

% (1,0) PU cell widefield images
\begin{figure*}[!th]
    \centering
    \includegraphics[width=\linewidth]{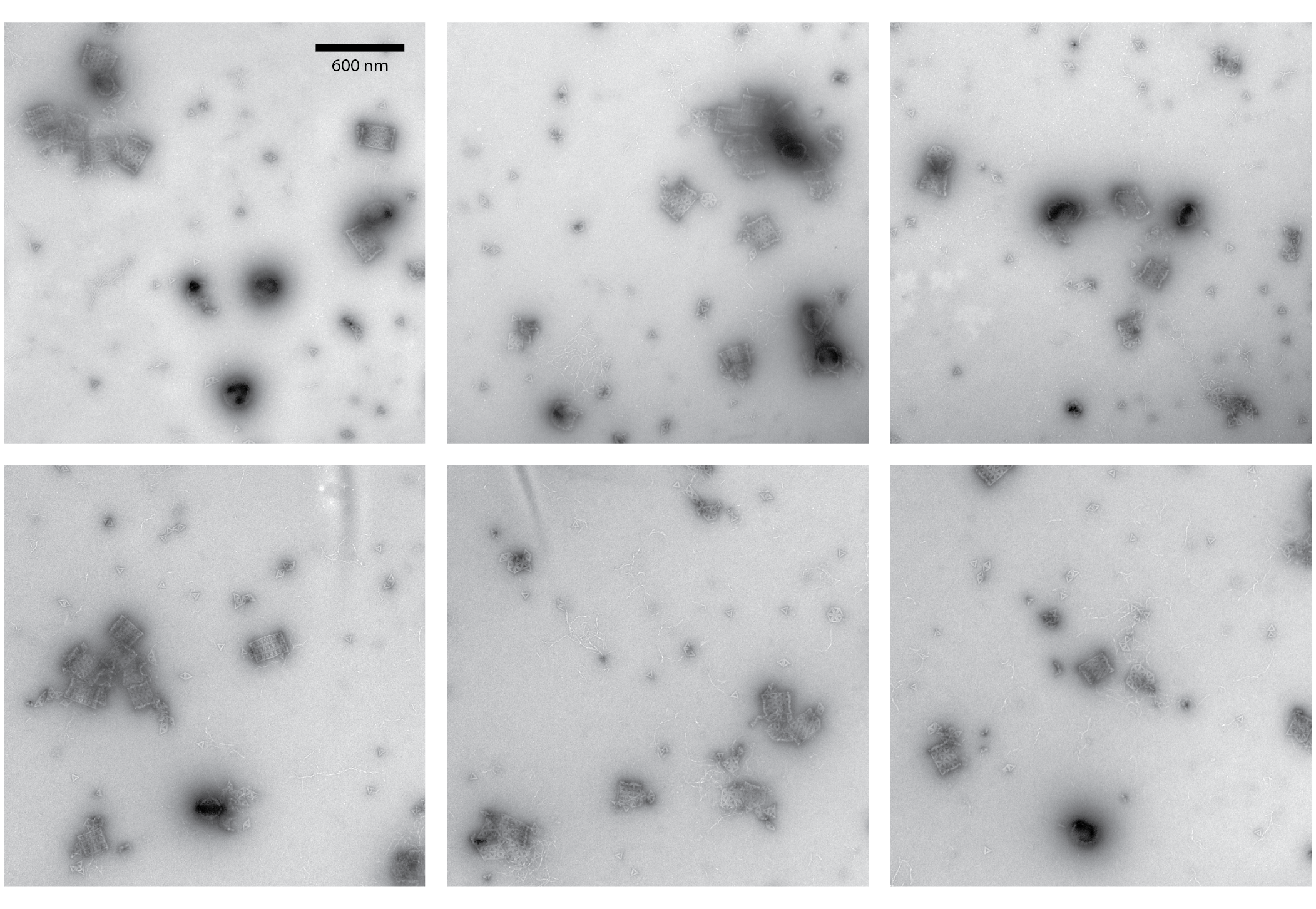}
    \caption{\textbf{Overview of (1,0) PU cell seeds.} The TEM images primarily show three-layer seeds with various widths, aggregates, unbound monomers, and oligomers. All images share the same scale bar.} 
    \label{SFig:(1,0)PU_cell_overview}
\end{figure*}

% (3,0) PU cell widefield images

\begin{figure*}[!th]
    \centering
    \includegraphics[width=\linewidth]{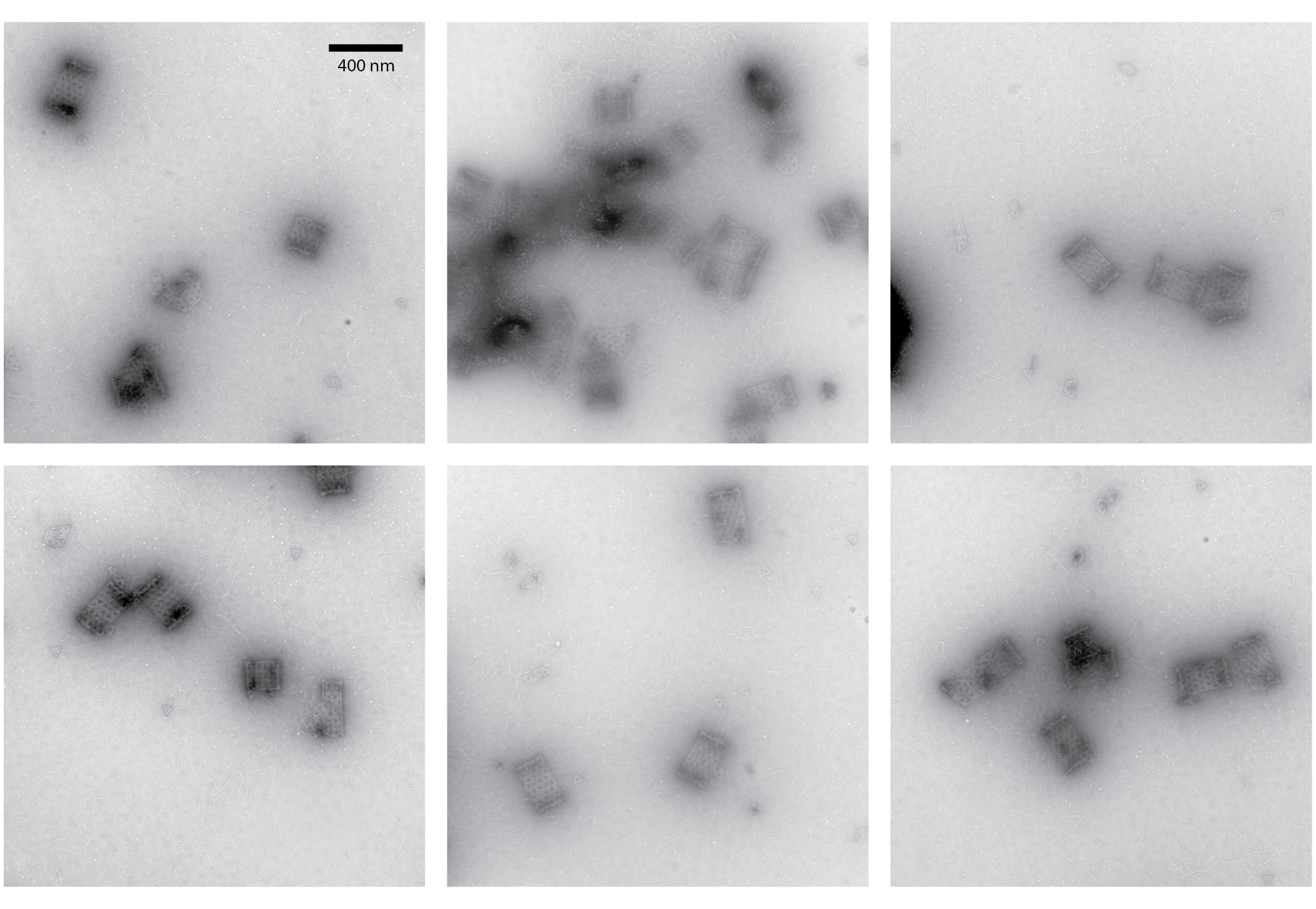}
    \caption{\textbf{Overview of (3,0) PU cell seeds.} The TEM images primarily show three-layer seeds with two dominant widths of (6,0) and (9,0) seeds, aggregates, unbound monomers, and oligomers. All images share the same scale bar.  
}
    \label{SFig:(3,0)PU_cell_overview}
\end{figure*}

% (4,0) PU cell widefield iamges

\begin{figure*}[!th]
    \centering
    \includegraphics[width=\linewidth]{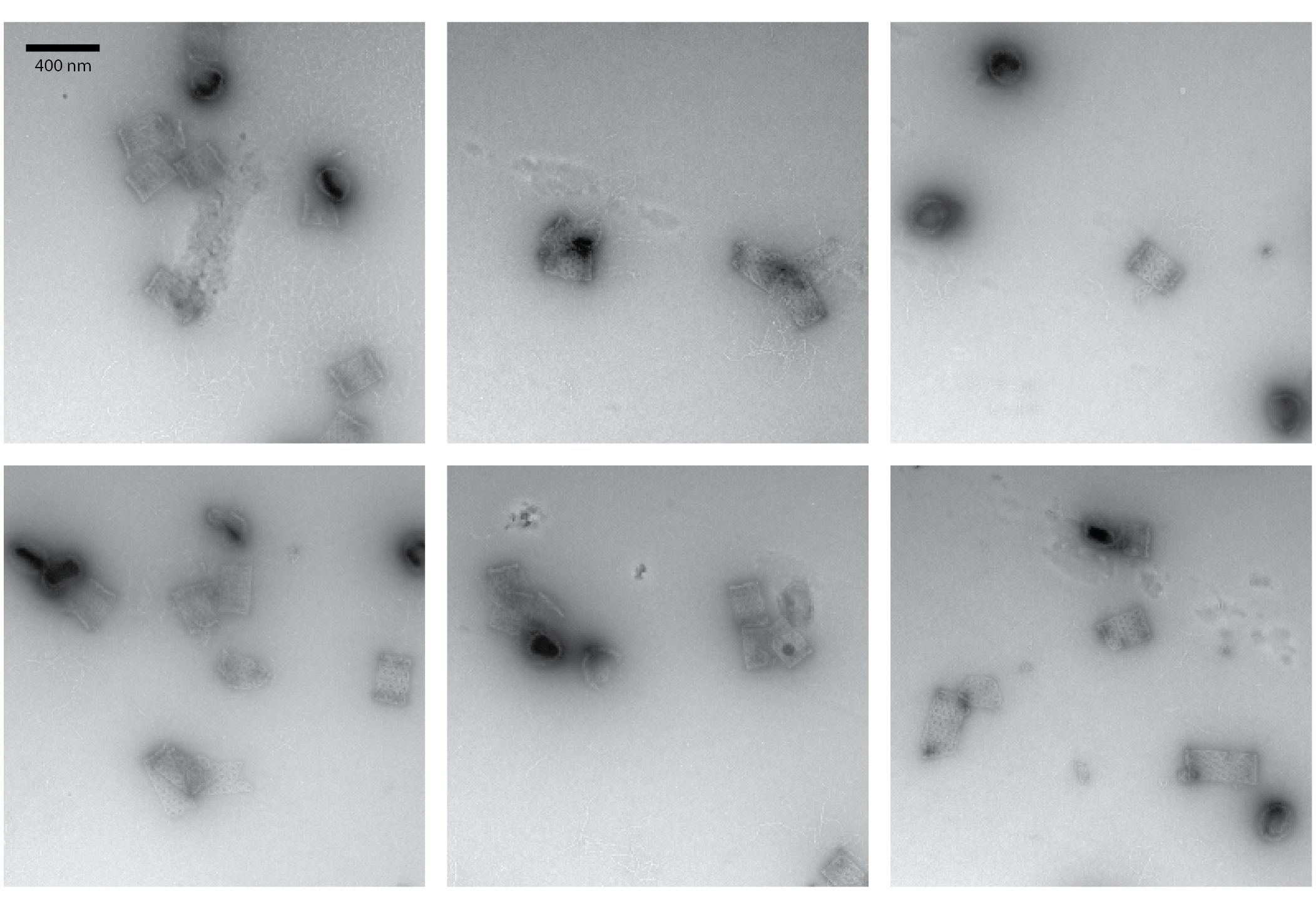}
    \caption{\textbf{Overview of (4,0) PU cell seeds.}  The TEM images primarily show three-layer seeds, which are dominated by (8,0) seeds, aggregates, unbound monomers, and oligomers. All images share the same scale bar.}
    \label{SFig:(4,0)PU_cell_overview}
\end{figure*}

% (7,0) UC widefield images
\begin{figure*}[!th]
    \centering
    \includegraphics[width=\linewidth]{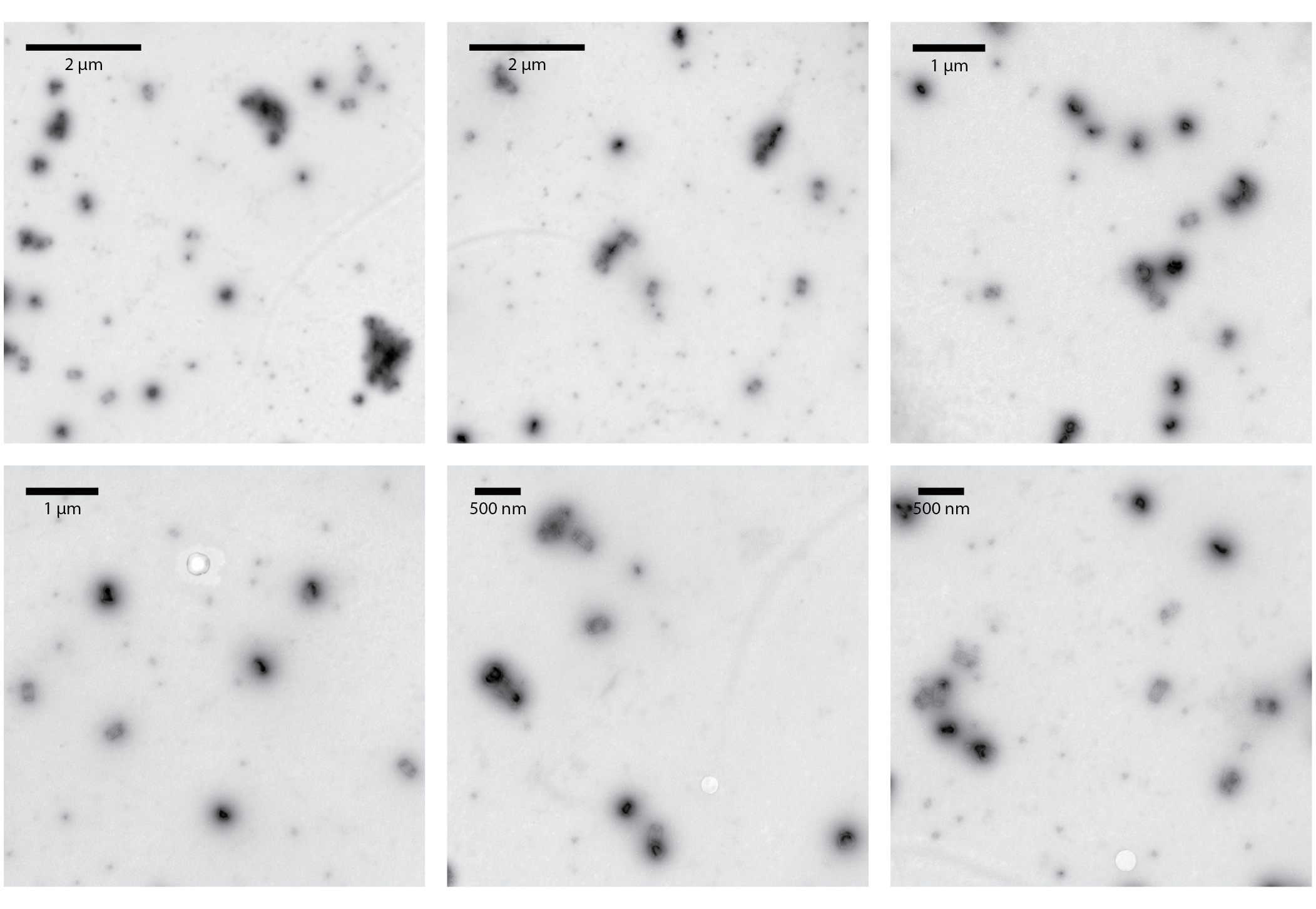}
    \caption{\textbf{Overview of the (7,0) PU cell seed.} The TEM images primarily display closed seeds, unclosed sheets, unbound monomers, oligomers, and aggregates. The black circular features correspond to seeds oriented vertically on the grid.  }
    \label{SFig:(7,0)PU_cell_overview}
\end{figure*}

% UV seed widefield image

\begin{figure*}[!th]
    \centering
    \includegraphics[width=\linewidth]{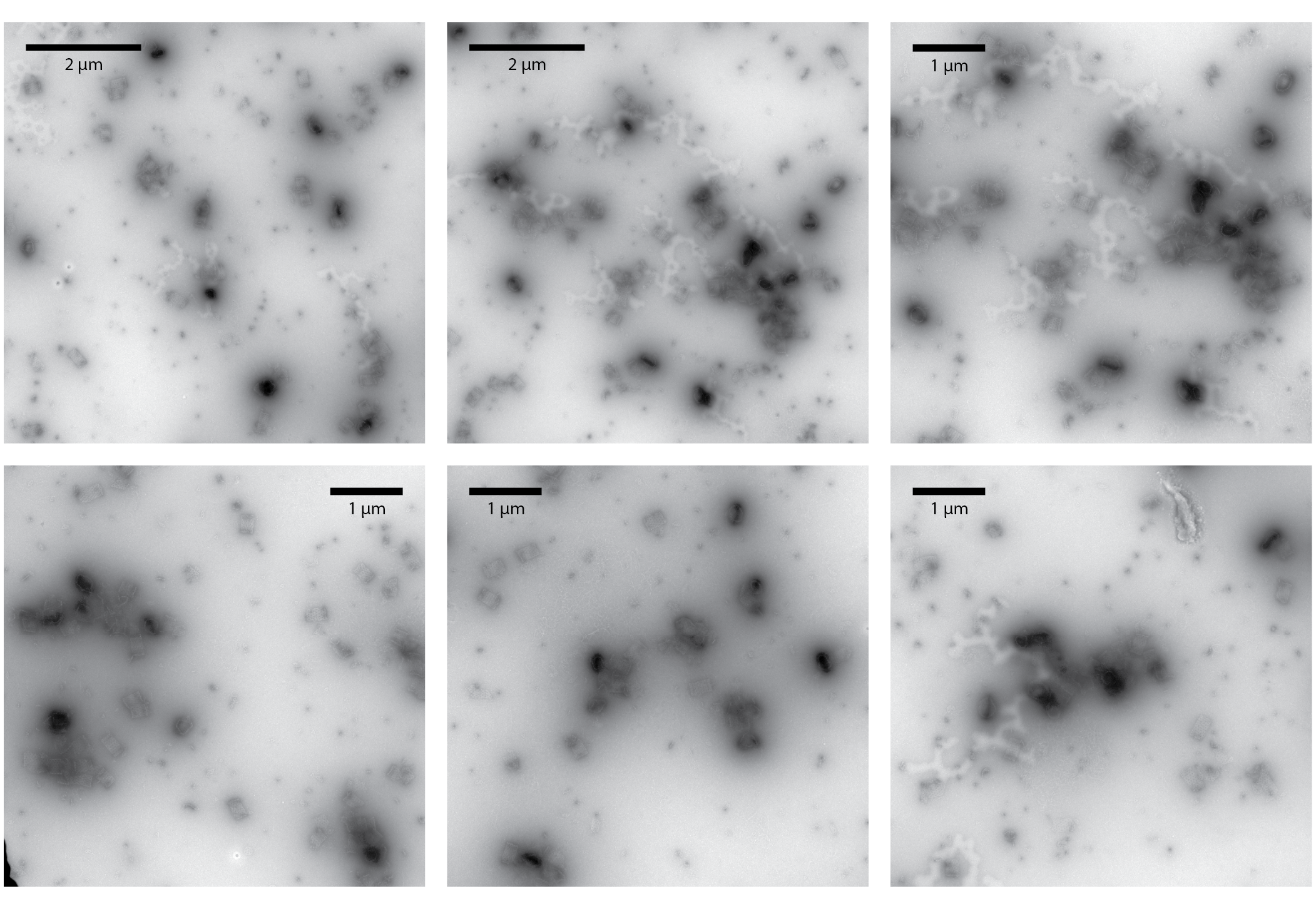}
    \caption{\textbf{Overview of UV weldable seed assembly.} The TEM images primarily show three-layer seeds with various widths, aggregates, unbound monomers, and oligomers. }
    \label{SFig:uv_seed}
\end{figure*}

\clearpage

%% 6 BASE PAIR SEQUENCE LIBRARY TABLE filtered %%
\begin{longtable}{p{0.15\linewidth} p{0.15\linewidth} p{0.15\linewidth} p{0.15\linewidth} p{0.15\linewidth} p{0.15\linewidth}}
\caption{\textbf{6 base pair interaction sequence library.} A list of 59 unique sequences, their complementary sequences, and an estimate of their binding free energy. } 
\label{tab:6bp_library}\\
Sequence & Compliment & $\Delta$G [kcal/mol] & Sequence & Compliment & $\Delta$G [kcal/mol]\\
\hline
\endfirsthead
Sequence & Compliment & $\Delta$G [kcal/mol] & Sequence & Compliment & $\Delta$G [kcal/mol] \\
\hline
\endhead

\endfoot
\endlastfoot
ACTAGC &	GCTAGT &    -6.09 & AGTTAC &	GTAACT &	-5.01\\
TAGTCT &	AGACTA &	-4.53 & CGATGG &	CCATCG &	-6.61\\
CCATTC &	GAATGG &	-5.61 & CTTGGT &	ACCAAG &	-5.92\\
ATGCAC &	GTGCAT &	-6.73 & AGTCAG &	CTGACT &	-5.85\\
CAATAG &	CTATTG &	-4.16 & TGATTG &	CAATCA &	-4.57\\
CTAGGA &	TCCTAG &	-4.77 & CACATC &	GATGTG &	-5.66\\
ACGAAG &	CTTCGT &	-6.29 & ACCTGA &	TCAGGT &	-5.93\\
ATGACA &	TGTCAT &	-5.14 & TACAGG & 	CCTGTA &	-5.08\\
AACCTA & 	TAGGTT &	-4.76 & GAGACA &	TGTCTC &	-5.29\\
GACAGA &	TCTGTC &	-5.29 & CGTCCA &	TGGACG &	-6.69\\
GCATCT &	AGATGC &	-6.09 & TATTCC &	GGAATA &	-4.26\\
AGATTC &	GAATCT &	-5.03 & TTCTCA &	TGAGAA &	-4.34\\
CTGTGA &	TCACAG &	-5.41 & TCGTAC &	GTACGA &	-5.59\\
AGTTCC & 	GGAACT &	-6.13 & CGATTA &	TAATCG &	-4.42\\
ATTCTG &	CAGAAT &	-5.01 & ATTCAG &	CTGAAT &	-5.01\\
CTTGAG &	CTCAAG &	-5.28 & GTAGAT & 	ATCTAC &	-4.42\\
GGATAA &	TTATCC &	-4.12 & TCATCC &	GGATGA &	-5.43\\
GGTATT &	AATACC &	-4.68 & GGTAAT &	ATTACC &	-4.68\\
ACTGAG &	CTCAGT &	-5.85 & AGAGAT &	ATCTCT &	-5.08\\
AGATAG &	CTATCT &	-4.42 & TTCCTG &	CAGGAA &	-5.36\\
TTCCAT &	ATGGAA &	-4.9 & GATATG &	CATATC &	-4.09\\
AACATT &	AATGTT &	-4.81 & TTGGCA &	TGCCAA &	-5.99\\
GACCTC &	GAGGTC &	-6.33 & CCTATG &	CATAGG &	-5\\
CTTAGG &	CCTAAG &	-4.95 & TAACAG &	CTGTTA &	-4.24\\
TCTTCT &	AGAAGA &	-4.59 & GTACAT &	ATGTAC &	-4.73\\
TTCAAT &	ATTGAA &	-4.06 & CTTACT &	AGTAAG &	-4.49\\
AGTATC &	GATACT &	-4.75 & GTATGT &	ACATAC &	-4.73\\
ACAATT &	AATTGT &	-4.81 & ACTAAC &	GTTAGT &	-5.01\\
CTTGTA &	TACAAG &	-4.24 & CTACAC &	GTGTAG &	-5.33\\
AACTAT &	ATAGTT &	-4.22
\end{longtable}

\clearpage

%% 6 base pair SIDE INTERACTION LIBRARY %% 
\begin{longtable}{p{0.11\linewidth} p{0.11\linewidth} p{0.11\linewidth} p{0.11\linewidth}p{0.11\linewidth} p{0.11\linewidth} p{0.11\linewidth} p{0.14\linewidth}}
\caption{\textbf{6 base pair side interactions for multicomponent assemblies.} A list of the set of six interaction sequences that make up a side interaction of a monomer and an estimate of their summed binding free energy. The first two sets of sequences are self-complementary, e.g. Position 1 binds to Position 6, Position 2 binds to Position 5, and Position 3 binds to Position 4. For the rest of the sets, X is complementary to X*.} 
\label{tab:6bp_side_interactions} \\
Name	&	Position 1	& 	Position 2	&	Position 3	&	Position 4	&	Position 5	&	Position 6	&	$\Delta$G [kcal/mol]	\\
\hline
\endfirsthead
Name	&	Position 1	& 	Position 2	&	Position 3	&	Position 4	&	Position 5	&	Position 6	&	$\Delta$G [kcal/mol]	\\
\hline
\endhead
sA side 1	&	ACTAGC	& 	AGTTAC	&	TAGTCT	&	AGACTA	&	GTAACT	&	GCTAGT	&	-30.57	\\
sA side 2	&	TTCAAT	& 	CCATTC	&	CTTGGT	&	ACCAAG	&	GAATGG	&	ATTGAA	&	-31.42	\\
A side 1	&	CAATAG	& 	TGATTG	&	CTAGGA	&	CACATC	&	ACGAAG	&	ACCTGA	&	-31.38	\\
A side 2	&	ATGACA	& 	TACAGG	&	AACCTA	&	GAGACA	&	GACAGA	&	ACTAAC	&	-30.57	\\
A side 3	&	GTACAT	& 	AGTCAG	&	CGATGG	&	CTTACT	&	AGTATC	&	GTATGT	&	-31.16	\\
A* side 1	&	TCAGGT	& 	CTTCGT	&	GATGTG	&	TCCTAG	&	CAATCA	&	CTATTG	&	-31.38	\\
A* side 2	&	GTTAGT	& 	TCTGTC	&	TGTCTC	&	TAGGTT	&	CCTGTA	&	TGTCAT	&	-30.57	\\
A* side 3	&	ACATAC	& 	GATACT	&	AGTAAG	&	CCATCG	&	CTGACT	&	ATGTAC	&	-31.16	\\
B side 1	&	GCATCT	& 	TATTCC	&	AGATTC	&	TTCTCA	&	CTGTGA	&	TCGTAC	&	-30.72	\\
B side 2	&	AGATAG	& 	TTCCTG	&	TTCCAT	&	GATATG	&	ATGCAC	&	AACATT	&	-30.31	\\
B side 3	&	ACAATT	& 	CGTCCA	&	CTTGTA	&	CTACAC	&	GACAGA	&	AACTAT	&	-30.58	\\
B* side 1	&	GTACGA	& 	TCACAG	&	TGAGAA	&	GAATCT	&	GGAATA	&	AGATGC	&	-30.72	\\
B* side 2	&	AATGTT	& 	GTGCAT	&	CATATC	&	ATGGAA	&	CAGGAA	&	CTATCT	&	-30.31	\\
B* side 3	&	ATAGTT	& 	TCTGTC	&	GTGTAG	&	TACAAG	&	TGGACG	&	AATTGT	&	-30.58	\\
C side 1	&	AGTTCC	& 	CGATTA	&	ATTCTG	&	ATTCAG	&	CTTGAG	&	GTAGAT	&	-30.27	\\
C side 2	&	GGATAA	& 	TCATCC	&	GGTATT	&	GGTAAT	&	ACTGAG	&	AGAGAT	&	-29.84	\\
C side 3	&	TTGGCA	& 	GACCTC	&	CCTATG	&	CTTAGG	&	TAACAG	&	TCTTCT	&	-31.1	\\
C* side 1	&	ATCTAC	& 	CTCAAG	&	CTGAAT	&	CAGAAT	&	TAATCG	&	GGAACT	&	-30.27	\\
C* side 2	&	ATCTCT	& 	CTCAGT	&	ATTACC	&	AATACC	&	GGATGA	&	TTATCC	&	-29.84	\\
C* side 3	&	AGAAGA	& 	CTGTTA	&	CCTAAG	&	CATAGG	&	GAGGTC	&	TGCCAA	&	-31.1	\\
D side 1	&	CACATC	& 	CAATAG	&	ACGAAG	&	TGATTG	&	ACCTGA	&	CTAGGA	&	-31.38	\\
D side 2	&	GAGACA	& 	ATGACA	&	GACAGA	&	TACAGG	&	ACTAAC	&	AACCTA	&	-30.57	\\
D side 3	&	CTTACT	& 	GTACAT	&	AGTATC	&	AGTCAG	&	GTATGT	&	CGATGG	&	-31.16	\\
D* side 1	&	TCCTAG	& 	TCAGGT	&	CAATCA	&	CTTCGT	&	CTATTG	&	GATGTG	&	-31.38	\\
D* side 2	&	TAGGTT	& 	GTTAGT	&	CCTGTA	&	TCTGTC	&	TGTCAT	&	TGTCTC	&	-30.57	\\
D* side 3	&	CCATCG	& 	ACATAC	&	CTGACT	&	GATACT	&	ATGTAC	&	AGTAAG	&	-31.16	\\

\end{longtable}

\clearpage

%% 7 BASE PAIR SEQUENCE LIBRARY TABLE %%
\begin{longtable}{p{0.15\linewidth} p{0.15\linewidth} p{0.15\linewidth}}
\caption{\textbf{7 base pair interaction library.} A list of 36 unique sequences, their complementary sequences, and an estimate of their binding free energy.} 
\label{tab:7bp_library}\\
Sequence & Compliment & $\Delta$G [kcal/mol] \\
\hline
\endfirsthead

Sequence & Compliment & $\Delta$G [kcal/mol]  \\
\hline
\endhead

\endfoot
\endlastfoot
AACAGTC & GACTGTT & -7.18 \\
TCCTGAC & GTCAGGA & -7.27 \\
CTGAGGA & TCCTCAG & -6.94 \\
TTGCTCA & TGAGCAA & -6.73 \\
TGAGTTG & CAACTCA & -6.41 \\
AGACAAT & ATTGTCT & -6.39 \\
ACACATC & GATGTGT & -7.23 \\
GAAGGTG & CACCTTC & -7.31 \\
GTGCTAG & CTAGCAC & -7.27 \\
TTAGGCA & TGCCTAA & -6.40 \\
GAGAACA & TGTTCTC & -6.29 \\
ATCACTC & GAGTGAT & -6.92 \\
CCAGTTC & GAACTGG & -7.45 \\
CCTACTG & CAGTAGG & -6.84 \\
GTCCAAT & ATTGGAC & -6.85 \\
GGATACT & AGTATCC & -6.26 \\
AAGAGAG & CTCTCTT & -6.54 \\
ACTAAGG & CCTTAGT & -6.52 \\
ATATGCA & TGCATAT & -6.10 \\
TCGTACT & AGTACGA & -6.64 \\
CCAAGTG & CACTTGG & -7.43 \\
TAGGTCC & GGACCTA & -6.94 \\
GTACCTG & CAGGTAC & -7.03 \\
GAATGAG & CTCATTC & -6.21 \\
CAATCCA & TGGATTG & -6.41 \\
CATCCTT & AAGGATG & -6.66 \\
GTAGACG & CGTCTAC & -7.21 \\
CGATCAA & TTGATCG & -6.59 \\
CCATACC & GGTATGG & -7.17 \\
AGCTTCT & AGAAGCT & -7.42 \\
GAAGATC & GATCTTC & -6.23 \\
ACCAGAT & ATCTGGT & -7.23 \\
CTTCAAC & GTTGAAG & -6.61 \\
GACCATT & AATGGTC & -6.85 \\
GATTCAC & GTGAATC & -6.54 \\
TCTCCAT & ATGGAGA & -6.48 \\

\end{longtable}

\clearpage

%% 7 BASE PAIR SIDE INTERACTION LIBRARY %%
\begin{longtable}{p{0.14\linewidth} p{0.11\linewidth} p{0.11\linewidth} p{0.11\linewidth}p{0.11\linewidth} p{0.11\linewidth} p{0.11\linewidth} p{0.11\linewidth}}
\caption{\textbf{7 base pair side interactions for multicomponent seed assemblies.} A list of the set of six interaction sequences that make up a side interaction of a monomer and an estimate of their summed binding free energy. The first nine sets of sequences are self-complementary, e.g. Position 1 binds to Position 6, Position 2 binds to Position 5, and Position 3 binds to Position 4. For the rest of the sets, X is complementary to X*. `sA reduced' is made by replacing position 2 and position 5 of the 7-base-pair sA interaction with position 2 and position 5 of the 6-base-pair sA interaction from Table~\ref{tab:6bp_side_interactions}.} 
\label{tab:7bp_side_interactions} \\
Name	&	Position 1	& 	Position 2	&	Position 3	&	Position 4	&	Position 5	&	Position 6	&	$\Delta$G [kcal/mol]	\\
\hline
\endfirsthead
Name	&	Position 1	& 	Position 2	&	Position 3	&	Position 4	&	Position 5	&	Position 6	&	$\Delta$G [kcal/mol]	\\
\hline
\endhead

sA side 1 & GAAGGTG & GACCATT & GAGAACA & TGTTCTC & AATGGTC & CACCTTC & -40.90 \\
sA side 2 & TTGCTCA & TGAGTTG & GTGCTAG & CTAGCAC & CAACTCA & TGAGCAA & -40.82 \\
sA side 3 & AACAGTC & AAGAGAG & AGACAAT & ATTGTCT & CTCTCTT & GACTGTT & -40.22 \\
sA reduced  side 1 & GAAGGTG & AGTTAC & GAGAACA & TGTTCTC & GTAACT & CACCTTC & -37.20 \\
sA reduced  side 2 & TTGCTCA & CCATTC & GTGCTAG & CTAGCAC & GAATGG & TGAGCAA & -39.20 \\
sA reduced  side 3 & AACAGTC & TCGACA & AGACAAT & ATTGTCT & TGTCGA & GACTGTT & -38.50 \\

sB side 1 & TTAGGCA & ACACATC & TCGTACT & AGTACGA & GATGTGT & TGCCTAA & -40.54 \\
sB side 2 & GAAGATC & CCTACTG & AGCTTCT & AGAAGCT & CAGTAGG & GATCTTC & -40.98 \\
sB side 3 & GGATACT & CTGAGGA & TCCTGAC & GTCAGGA & TCCTCAG & AGTATCC & -40.94 \\
C side 1 & ATATGCA & ATCACTC & CCAAGTG & TAGGTCC & GTACCTG & GAATGAG & -40.63\\
C side 2 & CAATCCA & CATCCTT & GTAGACG & CGATCAA & CCATACC & GTCCAAT & -40.89 \\
C side 3 & CCAGTTC & ACCAGAT & CTTCAAC & ACTAAGG & GATTCAC & TCTCCAT & -40.83 \\
C* side 1 & CTCATTC & CAGGTAC & GGACCTA & CACTTGG & GAGTGAT & TGCATAT & -40.63 \\
C* side 2 & ATTGGAC & GGTATGG & TTGATCG & CGTCTAC & AAGGATG & TGGATTG & -40.89 \\
C* side 3 & ATGGAGA & GTGAATC & CCTTAGT & GTTGAAG & ATCTGGT & GAACTGG & -40.83 \\
D side1 & TAGGTCC & ATATGCA & GTACCTG & ATCACTC & GAATGAG & CCAAGTG & -40.63 \\
D side2 & CGATCAA & CAATCCA & CCATACC & CATCCTT & GTCCAAT & GTAGACG & -40.89 \\
D side3 & ACTAAGG & CCAGTTC & GATTCAC & ACCAGAT & TCTCCAT & CTTCAAC & -40.83 \\
D* side1 & CACTTGG & CTCATTC & GAGTGAT & CAGGTAC & TGCATAT & GGACCTA & -40.63 \\
D* side2 & CGTCTAC & ATTGGAC & AAGGATG & GGTATGG & TGGATTG & TTGATCG & -40.89 \\
D* side3 & GTTGAAG & ATGGAGA & ATCTGGT & GTGAATC & GAACTGG & CCTTAGT & -40.83 \\
E side1 & GTACCTG & CCAAGTG & GAATGAG & ATATGCA & TAGGTCC & ATCACTC & -40.63 \\
E side2 & CCATACC & GTAGACG & GTCCAAT & CAATCCA & CGATCAA & CATCCTT & -40.89 \\
E side3 & GATTCAC & CTTCAAC & TCTCCAT & CCAGTTC & ACTAAGG & ACCAGAT & -40.83 \\
E* side1 & GAGTGAT & GGACCTA & TGCATAT & CTCATTC & CACTTGG & CAGGTAC & -40.63 \\
E* side2 & AAGGATG & TTGATCG & TGGATTG & ATTGGAC & CGTCTAC & GGTATGG & -40.89 \\
E* side3 & ATCTGGT & CCTTAGT & GAACTGG & ATGGAGA & GTTGAAG & GTGAATC & -40.83 \\
F side1 & TTGCTCA & GAAGGTG & TGAGTTG & GACCATT & GTGCTAG & GAGAACA & -40.86 \\
F side2 & TTAGGCA & AACAGTC & ACACATC & AAGAGAG & TCGTACT & AGACAAT & -40.38 \\
F side3 & GGATACT & GAAGATC & CTGAGGA & CCTACTG & TCCTGAC & AGCTTCT & -40.96 \\
F* side1 & TGTTCTC & CTAGCAC & AATGGTC & CAACTCA & CACCTTC & TGAGCAA & -40.86 \\
F* side2 & ATTGTCT & AGTACGA & CTCTCTT & GATGTGT & GACTGTT & TGCCTAA & -40.38 \\
F* side3 & AGAAGCT & GTCAGGA & CAGTAGG & TCCTCAG & GATCTTC & AGTATCC & -40.96 \\
G side1 & TGAGTTG & GAGAACA & GTGCTAG & GAAGGTG & TTGCTCA & GACCATT & -40.86 \\
G side2 & ACACATC & AGACAAT & TCGTACT & AACAGTC & TTAGGCA & AAGAGAG & -40.38 \\
G side3 & CTGAGGA & AGCTTCT & TCCTGAC & GAAGATC & GGATACT & CCTACTG & -40.96 \\
G* side1 & AATGGTC & TGAGCAA & CACCTTC & CTAGCAC & TGTTCTC & CAACTCA & -40.86 \\
G* side2 & CTCTCTT & TGCCTAA & GACTGTT & AGTACGA & ATTGTCT & GATGTGT & -40.38 \\
G* side3 & CAGTAGG & AGTATCC & GATCTTC & GTCAGGA & AGAAGCT & TCCTCAG & -40.96 \\
I side1 & GAAGGTG & TAGGTCC & TGAGTTG & ATCACTC & GAGAACA & GAATGAG & -40.08 \\
I side2 & AACAGTC & CGATCAA & ACACATC & CATCCTT & AGACAAT & GTCCAAT & -40.90 \\
I side3 & GAAGATC & ACTAAGG & CTGAGGA & ACCAGAT & AGCTTCT & TCTCCAT & -40.82 \\
I* side1 & CTCATTC & TGTTCTC & GAGTGAT & CAACTCA & GGACCTA & CACCTTC & -40.08 \\
I* side2 & ATTGGAC & ATTGTCT & AAGGATG & GATGTGT & TTGATCG & GACTGTT & -40.90 \\
I* side3 & ATGGAGA & AGAAGCT & ATCTGGT & TCCTCAG & CCTTAGT & GATCTTC & -40.82 \\
J side1 & ATATGCA & TTGCTCA & GTACCTG & GACCATT & CCAAGTG & GTGCTAG & -41.41 \\
J side2 & CAATCCA & TTAGGCA & CCATACC & AAGAGAG & GTAGACG & TCGTACT & -40.37 \\
J side3 & CCAGTTC & GGATACT & GATTCAC & CCTACTG & CTTCAAC & TCCTGAC & -40.97 \\
J* side1 & CTAGCAC & CACTTGG & AATGGTC & CAGGTAC & TGAGCAA & TGCATAT & -41.41 \\
J* side2 & AGTACGA & CGTCTAC & CTCTCTT & GGTATGG & TGCCTAA & TGGATTG & -40.37 \\
J* side3 & GTCAGGA & GTTGAAG & CAGTAGG & GTGAATC & AGTATCC & GAACTGG & -40.97 \\

\end{longtable}

%% 2 LAYERS SEED INTERACTIONS %%
\begin{longtable}{p{0.09\linewidth} p{0.09\linewidth} p{0.07\linewidth} p{0.07\linewidth}p{0.07\linewidth} p{0.05\linewidth} p{0.09\linewidth} p{0.09\linewidth} p{0.07\linewidth} p{0.07\linewidth}p{0.07\linewidth} }
\caption{\textbf{Subunit interactions for 2-layer seeds.} This table lists the interactions for each subunit for the (7,0) two-layer PU cell. The sequences are 7-base-pair interactions from Table~\ref{tab:7bp_side_interactions} unless otherwise indicated. If it is labeled `6bp-', as on some side 3 strands, those interactions come from Table~\ref{tab:6bp_side_interactions}. These seeds are complementary with the growing subunit `B' in Table~\ref{tab:tubule_interaction}.} \label{tab:2_layers_seed_interactoins} \\
	PU cell 	&	Subunit ID	&	Side 1 strands	&	Side 2 strands	&	Side 3 strands	\\
\hline
\endhead

\endfoot
\endlastfoot

 (7,0)  & 1 & F & C* & 6bp-C \\
        & 2 & C & C & C \\
        & 3 & F* & F* & C* \\
        & 4 & G & F & 6bp-C \\
        & 5 & C* & D* & 6bp-C \\
        & 6 & D & D & D \\
        & 7 & G* & G* & D* \\
        & 8 & I & G & 6bp-C \\
        & 9 & D* & E* & 6bp-C \\
        & 10 & E & E & E \\
        & 11 & I* & I* & E* \\
        & 12 & J & I & 6bp-C \\
        & 13 & E* & J & 6bp-C \\
        & 14 & J* & J* & sA \\
\end{longtable}

%% 3 LAYERS SEED INTERACTIONS %%
\begin{longtable}{p{0.09\linewidth} p{0.09\linewidth} p{0.07\linewidth} p{0.07\linewidth}p{0.07\linewidth} p{0.05\linewidth} p{0.09\linewidth} p{0.09\linewidth} p{0.07\linewidth} p{0.07\linewidth}p{0.07\linewidth} }
\caption{\textbf{Subunit interactions for 3-layers seed.} This table lists the interactions for each subunit for the (1,0), (3,0), and (4,0) three-layer PU cell. The sequences are 7-base-pair interactions from Table~\ref{tab:7bp_side_interactions} unless otherwise indicated. If it is labeled `6bp-', as on some side 3 strands, those interactions come from Table~\ref{tab:6bp_side_interactions}. These seeds are complementary with the growing particle `B' in Table~\ref{tab:tubule_interaction}.}
\label{tab:3_layers_seed_interactoins} \\
% Number of subunits	&	Subunit ID	&	Side 1 strands	&	Side 2 strands	&	Side 3 strands	&	&	Number of subunits	&	Subunit ID	&	Side 1 strands	&	Side 2 strands	&	Side 3 strands	\\
% \hline
% \endfirsthead

PU cell	&	Subunit ID	&	Side 1 strands	&	Side 2 strands	&	Side 3 strands	&	&	PU cell 	&	Subunit ID	&	Side 1 strands	&	Side 2 strands	&	Side 3 strands	\\
\hline
\endhead

\endfoot
\endlastfoot

(1,0)   & 1     & sA reduced    & sA reduced    & C     &  & (4,0)  & 1     & C     & C     & 6bp-C \\
        & 2     & C*            & C*            & C*    &  &        & 2     & D     & C*    & C* \\
        & 3     & C             & C             & 6bp-C &  &        & 3     & E     & D     & C \\
(3,0)   & 1     & C             & C             & 6bp-C &  &        & 4     & sA    & D*    & D* \\
        & 2     & D             & C*            & C*    &  &        & 5     & F     & E     & D \\
        & 3     & E             & D             & C     &  &        & 6     & D*    & E*    & 6bp-C \\
        & 4     & sA            & D*            & D*    &  &        & 7     & F*    & F     & 6bp-C \\
        & 5     & F             & E             & D     &  &        & 8     & G     & F*    & E \\
        & 6     & D*            & E*            & 6bp-C &  &        & 9     & E*    & G     & E* \\
        & 7     & F*            & F             & 6bp-C &  &        & 10    & G*    & H     & 6bp-C \\
        & 8     & C*            & F*            & E     &  &        & 11    & C*    & H*    & F \\
        & 9     & E*            & sA            & E*    &  &        & 12    & sB    & G*    & F* \\

\end{longtable}

%% UV seed interaction %%
\begin{longtable}{p{0.1\linewidth} p{0.1\linewidth} p{0.09\linewidth} p{0.09\linewidth}p{0.09\linewidth}}
\caption{\textbf{Subunit interactions for the three-layer UV-weldable seed.} This table lists the interaction for each subunit for the UV-weldable seed. The sequences are 6-base-pair interactions from Table~\ref{tab:6bp_side_interactions}. This seed is complementary with the growing particle X in Table~\ref{tab:tubule_interaction}. }
\label{tab:uv_seed_interactions}\\
PU cell	&	Subunit ID	&	Side 1 strands	&	Side 2 strands	&	Side 3 strands\\
\hline
\endhead

\endfoot
\endlastfoot

(1,0)   & 10    & B     & B     & A     \\
        & 11    & B*    & B*    & A     \\
        & 2     & A*    & A*    & A*    \\
        & 3     & A     & A     & D*     \\
\end{longtable}

%% Growing tubule interaction %%
\begin{longtable} {p{0.1\linewidth} p{0.09\linewidth} p{0.09\linewidth} p{0.09\linewidth}}

\caption{\textbf{Subunit interactions for the tubule monomers.} Particle A and B, or Particle X and Y, are complementary tubule monomers. In this study, we use A and B to assemble nucleated tubules and templated tubules grown from multispecies seed. We use X and Y to assemble templated tubules grown from the UV-weldable seed. The sequences are 6-base-pair interactions from Table~\ref{tab:6bp_side_interactions}.}

\label{tab:tubule_interaction}\\
Subunit ID	&	Side 1 strands	&	Side 2 strands	&	Side 3 strands\\
\hline
\endhead

\endfoot
\endlastfoot

    A       & C     & C     & C     \\
    B       & C*    & C*    & C*    \\
    X       & D     & D     & D     \\
    Y       & D*    & D*    & D*     \\

\end{longtable}

%\clearpage

\def\bibsection{\section*{Supplementary references}}